\documentclass[12pt,a4paper]{article}
\pdfoutput=1
\usepackage[utf8]{inputenc}
\usepackage{epsf,amsmath,amssymb,graphicx,dcolumn}
\usepackage{caption}
\usepackage{subcaption}
\usepackage{scalefnt,ulem,pstricks}
\usepackage{booktabs,multirow,tabularx}
\usepackage{cite,hyperref}
\usepackage{cleveref}
\usepackage{color}
\usepackage{rotating}
\usepackage{microtype}
\usepackage[titletoc,title]{appendix}
\usepackage[numbers,sort&compress]{natbib}
\usepackage{amsmath,amsfonts,amsthm,bm} % Math packages
\usepackage[tikz]{bclogo}
\usepackage{subcaption}
\usepackage{tikz-feynman}
 
\newcommand\F{${\rm F}$}
\newcommand\FJ{${\rm FJ}$}
\newcommand\FJJ{${\rm FJJ}$}
\newcommand\PhiB{\Phi_{\scriptscriptstyle \rm F}}

\newcommand\PhiBJ{\Phi_{\scriptscriptstyle \rm FJ}}

\newcommand{\pt}{{p_{\text{\scalefont{0.77}T}}}}
\newcommand{\ptrad}{{p_{\text{\scalefont{0.77}T,rad}}}}

\newcommand{\mh}{{m_{\text{\scalefont{0.77}H}}}}
\newcommand{\mz}{{m_{\text{\scalefont{0.77}Z}}}}
\newcommand{\mw}{{m_{\text{\scalefont{0.77}W}}}}
\newcommand{\mt}{{m_{\text{\scalefont{0.77}t}}}}
\newcommand{\muF}{{\mu_{\text{\scalefont{0.77}F}}}}
\newcommand{\muR}{{\mu_{\text{\scalefont{0.77}R}}}}
\newcommand{\KF}{K_{\text{\scalefont{0.77}F}}}
\newcommand{\KR}{K_{\text{\scalefont{0.77}R}}}

\newcommand{\noun}[1]{{\scshape #1}}

\newcommand{\POWHEG}{\noun{POWHEG}}

\newcommand{\POWHEGBOX}{\noun{POWHEG-BOX}}
\newcommand{\POWHEGBOXVtwo}{\noun{POWHEG-BOX-V2}}

\newcommand{\minlo}{{\noun{MiNLO$^{\prime}$}}}
\newcommand{\minnlo}{{\noun{MiNNLO$_{\rm PS}$}}}
\newcommand{\Matrix}{{\noun{Matrix}}}
\newcommand{\PYTHIA}[1]{\noun{Pythia{#1}}}

\newcommand{\abar}{\frac{\as}{2\pi}}
\newcommand{\abarmu}[1]{\frac{\as(#1)}{2\pi}}

\newcommand{\nnlops}{NNLO+PS}

\newcommand{\citere}[1]{Ref.\,\cite{#1}}

\newcommand{\eqn}[1]{Eq.\,(\ref{#1})}

\newcommand{\fig}[1]{Fig.\,\ref{#1}}

\newcommand{\sct}[1]{Section~\ref{#1}}

\newcommand{\app}[1]{Appendix~\ref{#1}}

\newcommand{\LambdaPWG}{\Lambda_{\rm pwg}}

\usepackage{xcolor}
\newcommand{\mathd}{\mathrm{d}}
\newcommand{\tmop}[1]{\ensuremath{\operatorname{#1}}}

\newcommand{\sphid}[1]{}

\providecommand{\href}[2]{#2}

\newcommand\as{\alpha_{\mathrm{S}}}

\def\to{\rightarrow}

\def\mt{m_t}
\def\mw{m_W}
\def\mz{m_Z}

\def\citere#1{\mbox{Ref.~\cite{#1}}}

\def\nnlo{{\it f}\,NNLO}

\makeatletter
\g@addto@macro\bfseries{\boldmath}
\makeatother

\begin{document} 
\begin{flushright}
\vspace*{-1.5cm}
CERN-TH-2020-084, LAPTH-027/20, MPP-2020-75\\
\end{flushright}
\vspace{0.cm}

\begin{center}
{\Large \bf M{\scalefont{0.76}I}NNLO$_{\text{PS}}$: Optimizing $2\to 1$ hadronic processes}\\[0.5cm]

\end{center}

\begin{center}
{\bf Pier Francesco Monni}$^{(a)}$, {\bf Emanuele Re}$^{(b)}$, and {\bf Marius Wiesemann$^{(c)}$}

$^{(a)}$ CERN, Theoretical Physics Department, CH-1211 Geneva 23, Switzerland\\
$^{(b)}$ LAPTh, Universit\'e Grenoble Alpes, USMB, CNRS, 74940 Annecy, France\\
$^{(c)}$ Max-Planck-Institut f\"ur Physik, F\"ohringer Ring 6, 80805
M\"unchen, Germany

\href{mailto:pier.monni@cern.ch}{\tt pier.monni@cern.ch}\\
\href{mailto:emanuele.re@lapth.cnrs.fr}{\tt emanuele.re@lapth.cnrs.fr}\\
\href{mailto:wieseman@mpp.mpg.de}{\tt marius.wiesemann@cern.ch}\\
\end{center}

\begin{center} {\bf Abstract} \end{center}\vspace{-1cm}
\begin{quote}
\pretolerance 10000

We consider the \minnlo{} method to consistently combine
next-to-next-to-leading order (NNLO) QCD calculations with
parton-shower simulations.  We identify the main sources of differences
between \minnlo{} and fixed-order NNLO predictions for inclusive
observables due to corrections beyond NNLO accuracy and present simple
prescriptions to either reduce or remove them.
Refined predictions are presented for Higgs, charged- and
neutral-current Drell Yan production.  The agreement with fixed-order
NNLO calculations is considerably improved for inclusive observables
and scale uncertainties are reduced.
The codes are released within the \POWHEGBOX.

\end{quote}

\parskip = 1.2ex 

\section{Introduction}

Precision studies play a crucial role in the rich physics programme at
the Large Hadron Collider (LHC).  Not only do they enable the accurate
determination of Standard-Model (SM) rates and parameters, but they
also provide a valuable route to the discovery of new-physics
phenomena through small deviations from the SM.  Experimental
analyses rely on parton-shower simulations to generate fully exclusive
events.  Therefore, in order to fully exploit the vast amount
high-quality data collected at the LHC, it is now paramount to include
highest-order perturbative information in event simulation.

The consistent combination of next-to-next-to-leading order (NNLO) QCD
calculations with parton-shower simulations (\nnlops{}) is one of the
current challenges in collider theory, and it is indispensable to
provide the interface between accurate theory predictions and
precision measurements.  Four NNLO+PS
methods~\cite{Hamilton:2012rf,Alioli:2013hqa,Hoeche:2014aia,Monni:2019whf},
which rely on different theoretical formulations, have been proposed
in the past decade. A good \nnlops{} method should attain NNLO
accuracy for observables inclusive in the QCD radiation beyond the
Born level, while preserving the logarithmic structure (and accuracy)
of the parton-shower simulation after matching. While NNLO accuracy is
guaranteed by all existing methods, the kinematic constraints that
each of the above methods impose on the subsequent parton-shower
evolution may have consequences in terms of the logarithmic accuracy
of the final simulation. In Ref.~\cite{Monni:2019whf} we have
presented the method \minnlo{}, which has the following features:
\begin{itemize}
\item NNLO corrections are calculated directly during the generation
  of the events and without additional reweighting.
\item No merging scale is required to separate different multiplicities in the
  generated event samples.
\item The matching to the parton shower is performed according to the
  \POWHEG{} method~\cite{Nason:2004rx} and preserves the leading 
  logarithmic (LL) structure for transverse-momentum ordered 
  showers.\footnote{For a different ordering variable, preserving the
    accuracy of the shower is more subtle. Not only one needs to veto
    shower radiation that has relative transverse momentum greater
    than the one generated by \POWHEG{}, but also one has to resort to
    truncated showers~\cite{Nason:2004rx,Bahr:2008pv} to compensate
    for missing collinear and soft radiation. Failing to do so spoils the
    shower accuracy at leading-logarithmic level (in fact, at the
    double-logarithmic level).}
\end{itemize}

In this article we investigate the sources of differences between \minnlo{} and fixed-order NNLO (\nnlo{}) QCD predictions due to higher-order corrections
beyond the nominal perturbative accuracy. These differences affect
inclusive observables such as the total cross section or the rapidity
distribution of a color-singlet produced in hadronic collisions.
We identify the main sources of such corrections, which stem from:
\begin{enumerate}
\item Presence of higher-order terms (i.e. beyond NNLO) in the
  matching formula;
\item Scale setting in the QCD running coupling and in the
  parton distribution functions (PDFs);
\item Higher-order effects due to the parton shower recoil scheme.
\end{enumerate}
We introduce various prescriptions to either remove or reduce these corrections.
This leads to a significantly improved agreement between \minnlo{}
predictions and \nnlo{} calculations for inclusive
observables.
As a case study we focus on $2\to 1$ processes at the LHC, including Higgs boson production
as well as charged-current and neutral-current Drell Yan (DY) production, and we
present updated predictions that supersede those given in
Ref.~\cite{Monni:2019whf}. The computer codes with the implementation of
the \minnlo{} method for $2\to 1$ processes is released with this
article within the
\POWHEGBOX{} framework~\cite{Nason:2004rx,Frixione:2007vw,Alioli:2010xd}.
  
\section{M{\scalefont{0.76}I}NNLO$_{\text{PS}}$ in a nutshell}
\label{sec:minnlo}
The \minnlo~method~\cite{Monni:2019whf} formulates a NNLO calculation
fully differential in the phase space $\PhiB$ of the produced colour 
singlet \F{} with invariant mass $Q$. It starts from a differential description of the
production of the colour singlet and a jet (\FJ{}),
whose phase space we denote by $\PhiBJ$:\footnote{We note that this equation 
corresponds precisely to the one of a \POWHEG{} calculation for \FJ{} production, but with 
a modified content of the ${\bar B}(\PhiBJ)$ function.}
\begin{align}
\frac{\mathd\sigma}{\mathd\PhiBJ}={\bar B}(\PhiBJ) \times
\bigg\{\Delta_{\rm pwg} (\LambdaPWG) + \int\mathd \Phi_{\tmop{rad}} 
  \Delta_{\rm pwg} (\ptrad)  \frac{R (\PhiBJ{}, \Phi_{\tmop{rad}})}{B
  (\PhiBJ{})}\bigg\}\,,
\label{eq:master}
\end{align}
where ${\bar B}(\PhiBJ)$ generates the first radiation, while the
content of the curly brackets describes the generation of the second
radiation according to the
\POWHEG{}~method~\cite{Nason:2004rx,Frixione:2007vw,Alioli:2010xd}.
Here, $B$ and $R$ are the squared tree-level matrix elements for \FJ{}
and \FJJ{} production, respectively.  $\Delta_{\rm pwg}$ denotes the
\POWHEG{} Sudakov form factor~\cite{Nason:2004rx} and
$ \Phi_{\tmop{rad}} $ ($\ptrad$) is the phase space (transverse
momentum) of the second radiation. The \POWHEG{} cutoff $\LambdaPWG$
is used in the generation of the second radiation and its default
value is $\LambdaPWG=0.89$~GeV. The parton shower then
adds additional radiation to Eq.~\eqref{eq:master} that contributes
beyond $\mathcal{O}(\as^2(Q))$ at all orders in perturbation theory.
We refer to the explicit formulae of the original
publications~\cite{Nason:2004rx,Frixione:2007vw,Alioli:2010xd}.

The function ${\bar B}(\PhiBJ)$ is the central ingredient of
\minnlo{}. Its derivation~\cite{Monni:2019whf} stems from the
observation that the NNLO cross section differential in the transverse
momentum of the color singlet ($\pt$) and in the Born phase space
$\PhiB$ is described by the following formula
\begin{align}
\label{eq:start}
  \frac{\mathd\sigma}{\mathd\PhiB\mathd \pt} &= \frac{\mathd}{\mathd \pt}
     \bigg\{ \exp[-\tilde{S}(\pt)] {\cal L}(\pt)\Bigg\} +
                                               R_f(\pt) =
  \exp[-\tilde{S}(\pt)]\left\{
                                  D(\pt)+\frac{R_f(\pt)}{\exp[-\tilde{S}(\pt)]}\right\}\,,
\end{align}
where $R_f$ contains terms that are non-singular in the $\pt\rightarrow 0$ limit, and 
\begin{equation}
\label{eq:Dterms}
  D(\pt)  \equiv -\frac{\mathd \tilde{S}(\pt)}{\mathd \pt} {\cal L}(\pt)+\frac{\mathd {\cal L}(\pt)}{\mathd \pt}\,.
\end{equation}
$\tilde{S}(\pt)$, defined in Eq.~\eqref{eq:Rdef}, represents the Sudakov form factor, while ${\cal L}(\pt)$
contains the parton luminosities, the squared virtual matrix elements
for the underlying \F{} production process up to two loops as well as
the NNLO collinear coefficient functions and is given in \eqn{eq:luminosity} in
\app{app:formulae} (see Ref.~\cite{Monni:2019whf} for further
details). 
A crucial feature of the \minnlo{} method is that the renormalisation
and factorisation scales are set to $\muR\sim\muF\sim \pt$.

We introduce the NLO differential cross section for \FJ{} production 
\begin{equation}
\label{eq:NLO}
\frac{\mathd\sigma^{\rm (NLO)}_{\scriptscriptstyle\rm FJ}}{\mathd\PhiB\mathd
      \pt} = \abarmu{\pt}\left[\frac{\mathd\sigma_{\scriptscriptstyle\rm FJ}}{\mathd\PhiB\mathd
      \pt}\right]^{(1)} + \left(\abarmu{\pt}\right)^2\left[\frac{\mathd\sigma_{\scriptscriptstyle\rm FJ}}{\mathd\PhiB\mathd
      \pt}\right]^{(2)}\,,
\end{equation}
where $[X]^{(i)}$ denotes the coefficient of the
$i$-th term in the perturbative expansion of the quantity $X$, which
allows us to rewrite Eq.~\eqref{eq:start} as
\begin{align}
\label{eq:minnlo}
  \frac{\mathd\sigma}{\mathd\PhiB\mathd \pt}  &=
  \exp[-\tilde{S}(\pt)]\bigg\{ \abarmu{\pt}\left[\frac{\mathd\sigma_{\scriptscriptstyle\rm FJ}}{\mathd\PhiB\mathd
      \pt}\right]^{(1)} \left(1+\abarmu{\pt} [\tilde{S}(\pt)]^{(1)}\right)
  \notag
+ \left(\abarmu{\pt}\right)^2\left[\frac{\mathd\sigma_{\scriptscriptstyle\rm FJ}}{\mathd\PhiB\mathd
      \pt}\right]^{(2)} \notag\\
& + \left[D(\pt) -\abarmu{\pt} [D(\pt)]^{(1)}
  -\left(\abarmu{\pt}\right)^2 [D(\pt)]^{(2)}  \right]+ {\rm
  regular~terms~of~{\cal O}(\as^3)}\bigg\},
\end{align}
where the expressions of the $[D(\pt)]^{(i)}$ coefficients are given
in \app{app:formulae}.
The NNLO fully differential cross section is then obtained upon
integration over $\pt$ from scales of the order of the Landau pole
$\Lambda$ to the kinematic upper bound (we will discuss how to deal
with the Landau divergence and integrate down to arbitrarily small
$\pt$ in Section~\ref{sec:scales}). Each term of Eq.~\eqref{eq:minnlo}
contributes to the total cross section with scales
$\muR\sim\muF\sim Q$ according to the power counting formula
\begin{equation}
    \int_{\Lambda}^{Q} \mathd \pt \frac{1}{\pt} \as^m(\pt) \ln^n\frac{Q}{\pt}
\exp(-\tilde{S}(\pt))    \approx {\cal O}\left(\as^{m-\frac{n+1}{2}}(Q)\right)\,.
\end{equation}
This suggests that one can expand the second line of 
Eq.~\eqref{eq:minnlo}, while neglecting terms that, upon integration over
$\pt$, produce N$^3$LO corrections or beyond to any inclusive observable in $\PhiB$. 
We can therefore truncate the second line of 
Eq.~\eqref{eq:minnlo} to third order in $\as(\pt)$
\begin{equation}
\label{eq:truncated_D}
D(\pt) -\abarmu{\pt} [D(\pt)]^{(1)} -\left(\abarmu{\pt}\right)^2
[D(\pt)]^{(2)} =\left(\abarmu{\pt}\right)^3 [D(\pt)]^{(3)} +{\cal O}(\as^4(\pt))\,.
\end{equation}
The above considerations can be made at the fully differential level
on the $\PhiBJ$ phase space,  which leads to the definition of the
${\bar B}(\PhiBJ)$ function as~\cite{Monni:2019whf}
\begin{align}
\label{eq:Bbar}
{\bar B}(\PhiBJ)&\equiv \exp[-\tilde{S}(\pt)]\bigg\{ \abarmu{\pt}\left[\frac{\mathd\sigma_{\scriptscriptstyle\rm FJ}}{\mathd\PhiBJ}\right]^{(1)} \left(1+\abarmu{\pt} [\tilde{S}(\pt)]^{(1)}\right)\notag
  \\
&+ \left(\abarmu{\pt}\right)^2\left[\frac{\mathd\sigma_{\scriptscriptstyle\rm FJ}}{\mathd\PhiBJ}\right]^{(2)} + \left(\abarmu{\pt}\right)^3 [D(\pt)]^{(3)}  F^{\tmop{corr}}(\PhiBJ)\bigg\}\,,
\end{align}
where the factor $F^{\tmop{corr}}(\PhiBJ)$ encodes the dependence of
the correction $[D(\pt)]^{(3)}$ upon the full $\PhiBJ$ phase space, as discussed 
in detail in Section 3 of \citere{Monni:2019whf}.

\section{Implementation and corrections beyond NNLO}
\label{sec:implementation}
The derivation of ${\bar B}(\PhiBJ)$ in Eq.~\eqref{eq:Bbar} relies on
the fact that the running coupling and the parton densities are
evaluated at scales $\muR\sim\muF\sim \pt$. This is crucial to ensure
that we only neglect corrections that give rise to N$^3$LO terms or beyond in
the integrated cross section when truncating the expression in curly brackets
in the ${\bar B}(\PhiBJ)$ function at the third order in $\as(\pt)$ .
This procedure introduces a sensitivity of observables inclusive over QCD radiation 
to the small-$\pt$ region. Specifically, one has to ensure
that, when integrating over $\pt$, ${\bar B}(\PhiBJ)$ is
evaluated accurately down to sufficiently low $\pt$, until
the Sudakov suppression makes it tend to zero exponentially.

In practice, one meets the following problems:
\begin{enumerate}
\item The approximation in \eqn{eq:truncated_D}, while formally
  correct, introduces a treatment of subleading corrections quite different 
  from \nnlo{} calculations that might lead to numerically sizeable 
  differences in specific processes and in configurations
  where the $\pt$ of the colour singlet is small. By avoiding the truncation of the series 
  done in~\eqn{eq:truncated_D} one may thus reduce the contamination from higher-order 
  corrections with respect to fixed order. 

\item The parton densities are extracted from fits at a low scale
  $\Lambda_{\rm PDF}$ of the order of the proton mass and are effectively
  frozen or cut off at this scale. Moreover, some PDF sets contain an
  intrinsic charm component that requires $\Lambda_{\rm PDF}$ to be
  above the charm mass. 
  In a \minnlo{} calculation such scales are potentially too high for certain processes, 
  as one becomes
  sensitive to the PDF cutoff for  $\pt\sim\Lambda_{\rm PDF}$
  ($\pt \sim 2\, \Lambda_{\rm PDF}$ when scale variation is
  performed). 
  For DY production at $Q\sim M_V$ for instance, where $M_V$ is
  the invariant mass of the vector boson,
  these scales are dangerously close to the peak of the $\pt$
  distribution, and freezing the PDFs in~\eqn{eq:Bbar} at $\Lambda_{\rm PDF}$ 
  may cause undesired artefacts in some phase space regions.
  One therefore needs a prescription to carry out the PDFs evolution
  down to lower scales consistently.

\item One essential element of parton-shower algorithms is the recoil
  scheme, i.e.\ the choice of how the kinematic recoil of a new
  emission is distributed among the other particles in the event. In
  many schemes also the kinematics of the colour singlet is affected
  by the shower radiation. Thus, the parton shower may change
  inclusive observables in regions sensitive to infrared physics, for
  instance when the singlet is produced with large absolute
  rapidity. One may reduce these effects by choosing a recoil scheme
  that affects less the kinematics of the colour singlet.
\end{enumerate}

In the following, we will address each of the above points in more detail.

\subsection{Higher-order differences between M{\scalefont{0.76}I}NNLO$_{\text{PS}}$ and \nnlo{}}
\minnlo{} and \nnlo{} calculations differ by terms beyond their nominal accuracy.
The integration of Eq.~\eqref{eq:Bbar} over $\pt$ reproduces the fully
differential cross section up to $\mathcal{O}(\as^2)$. 
However, this result is affected by the truncation of the
$D(\pt)$ function in \eqn{eq:truncated_D}. This can be easily understood
by noticing that after this truncation the integral does not reproduce the exact
total derivative that we started with in \eqn{eq:start}. The truncated terms, although formally subleading,
can be numerically relevant in configurations in which $\pt$ is
small. The total derivative can be restored by avoiding 
the approximation of \eqn{eq:truncated_D}. Therefore, we retain the
option to generate events without truncating the $D(\pt)$ function
 by replacing in Eq.~\eqref{eq:Bbar}
\begin{align}
\label{eq:Dterms_new}
 \left(\abarmu{\pt}\right)^3[D(\pt)]^{(3)}  &\to -\frac{\mathd \tilde{S}(\pt)}{\mathd \pt} {\cal L}(\pt)+\frac{\mathd {\cal L}(\pt)}{\mathd \pt} -\abarmu{\pt} [D(\pt)]^{(1)}- \left(\abarmu{\pt}\right)^2[D(\pt)]^{(2)}\,,
\end{align}
where all ingredients are given in \app{app:formulae}.
In practice this is done by evaluating the full luminosity factor
${\cal L}(\pt)$ (see Section 4 of
Ref.~\cite{Monni:2019whf} for its derivation), and by performing its
derivative numerically for each event.  This prescription considerably
reduces the difference between \minnlo{} and \nnlo{} calculations, with the latter having
perturbative scales of the order of the invariant mass of the
colour singlet. These differences are strictly due to higher-order
corrections beyond NNLO, and the goal of the prescription in
\eqn{eq:Dterms_new} is to eliminate the main source of such subleading
terms.
This does not mean, however, that the integration of
Eq.~\eqref{eq:Bbar} reproduces the \nnlo{} total cross section at
$\muR=\muF=Q$, as the scale setting in the \minnlo{} approach is
rather different and fixed by the structure of $\pt$ resummation.

\begin{figure}[t]
  \centering
  \includegraphics[width=0.6\textwidth]{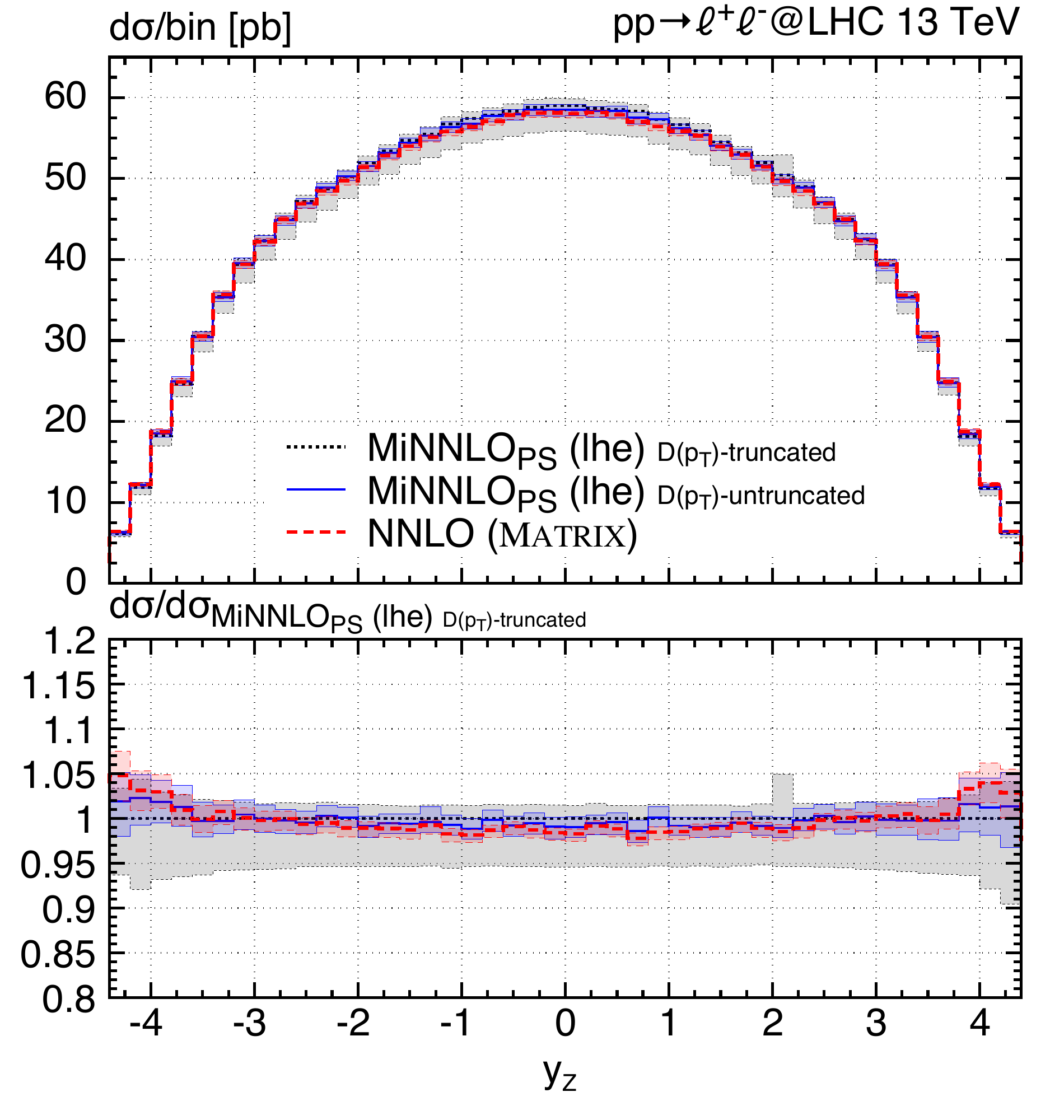}
  \caption{Rapidity distribution of the $Z$ boson. The plot compares
    the \minnlo{} prediction with untruncated (blue solid) and
    truncated (black dotted) $D(\pt)$ function at the LHE level with
    the NNLO prediction obtained with \Matrix{} (red, dashed).
    The lower panel shows the ratio to the truncated prediction.}
  \label{fig:truncation_comparisons}
\end{figure}

It is instructive to quantify the effect of this change. As a case
study we consider the rapidity distribution of the $Z$ boson in the
setup detailed in Section~\ref{sec:results}. To this end,
\fig{fig:truncation_comparisons} compares \minnlo{} predictions with
untruncated, using Eq.~\eqref{eq:Dterms_new}, and truncated, using
Eq.~\eqref{eq:truncated_D}, $D(\pt)$ function at the Les Houches Event
(LHE) level with the \nnlo{} prediction obtained with
\Matrix{}~\cite{Grazzini:2017mhc}.  We clearly observe an improvement
in the agreement between \minnlo{} and \nnlo{} when
using the untruncated prescription, both at the level of the shape of
the distribution and (even more notably) at the level of the
perturbative scale uncertainties, which are now comparable between the
two calculations. We recall \cite{Monni:2019whf} that \minnlo{}
includes an additional scale variation in the Sudakov form factor,
which has a mild effect. This provides a more reliable estimate of the
perturbative uncertainties associated with the matching procedure.
The improvement in the scale dependence can be understood by noticing
that the terms neglected in Eq.~\eqref{eq:truncated_D} are (although
formally subleading) logarithmically enhanced and therefore become
important around the Sudakov peak of the $\pt$ distribution where the
bulk of the cross section originates from. The inclusion of such terms
through the prescription of \eqn{eq:Dterms_new}  eliminates this feature and
results in a more reliable uncertainty band.
We notice a difference between the truncated results of
Fig.~\ref{fig:truncation_comparisons} and the corresponding
distribution shown in Ref.~\cite{Monni:2019whf} in the size
of the uncertainty band. This is mainly due to the different PDF set
(NNPDF3.1~\cite{Ball:2017nwa}) used in this article that comes with
a higher cutoff scale as well as the improved treatment of the
PDF evolution adopted here, which is discussed in the next section.

We employ the untruncated prescription of
\eqn{eq:Dterms_new} as the new default in the \minnlo{} method and in all 
results shown in the following.

% \begin{figure}[!htpb]

\subsection{Evolution of parton densities and scale setting}
\label{sec:scales}
As a second aspect that affects \minnlo{} predictions we discuss
the evolution of parton densities at low
scales. To ensure consistency of $\muF$ variations
we would like to avoid
truncating the PDFs at their own infrared
cutoff $\Lambda_{\rm PDF}$, but rather carry out a consistent DGLAP
evolution down to lower scales.
By doing this, we do not aim for a physically accurate
description of the $\pt$ spectrum below $\Lambda_{\rm PDF}$, but
simply ensure that Eq.~\eqref{eq:Bbar} can be evaluated all the way
down to sufficiently small $\pt$, where it becomes vanishingly
small due to the Sudakov suppression. This kinematic region will subsequently be corrected by the
parton shower and hadronization process.
As a consequence, several prescriptions can be formulated.

We read the PDFs (including the corresponding heavy quark thresholds)
from the {\tt LHAPDF}~\cite{Buckley:2014ana} package and build
corresponding {\tt HOPPET} grids~\cite{Salam:2008qg}, which
facilitates an efficient evaluation of all convolutions with the
coefficient functions. These {\tt HOPPET} grids are a copy of the {\tt
  LHAPDF} sets for $\muF \gtrsim \Lambda_{\rm PDF}$. Below this scale,
we freeze the number of active flavours to those of the PDF set at
$\muF =\Lambda_{\rm PDF}$, and we carry out a DGLAP evolution down to
lower scales of $\muF\sim \Lambda$ using {\tt HOPPET}.  This
prescription allows us to define a hybrid PDF set that can be
evaluated consistently also for small values of $\muF\sim \pt$ with
the desired numerical precision.
For consistency, we adopt the same running coupling as that provided
by the PDF set via {\tt LHAPDF}, with the full heavy quark threshold
information. The integral that defines the Sudakov form factor
$\tilde{S}(\pt)$ given in Eq.~\eqref{eq:Rdef} is then evaluated
exactly in numerical form. This numerical prescription replaces the
analytic formulae of $\tilde{S}(\pt)$  given in Ref.~\cite{Monni:2019whf}.

In the practical implementation of \citere{Monni:2019whf} we followed
a prescription to smoothly turn off the contribution of $[D(p_T)]^{(3)}$ in
Eq.~\eqref{eq:Bbar} at large $\pt$ by introducing modified logarithms
\begin{equation}
\label{eq:modlog}
\ln\frac{Q}{\pt} \to L\equiv \frac{1}{p}\ln\left(1+\left(\frac{Q}{\pt}\right)^p\right)\,,
\end{equation}
where $p$ is a free positive parameter. Larger values of $p$
correspond to logarithms that tend to zero at a faster rate at large
$\pt$, while the limit $\pt\to 0$ remains unaffected.  
This prescription modifies \eqn{eq:Bbar} by terms beyond accuracy, and
it has to be performed at the level of Eq.~\eqref{eq:start} in order
to preserve the total derivative (hence the total cross section).
This corresponds to:
\begin{itemize}
\item Setting the perturbative scales in the $D(\pt)$ (or $[D(p_T)]^{(3)}$)
  function of Eq.~\eqref{eq:Bbar} to
\begin{equation}
\label{eq:scales}
\muR = \KR \,Q \,e^{-L}\,,\qquad \muF = \KF \,Q\, e^{-L}\,.
\end{equation}
\item Changing the lower integration bound of the Sudakov~\eqref{eq:Rdef} (the
  integrand is not modified directly) to
\begin{equation}
\pt \to Q e^{-L}.
\end{equation}
\item Multiply $D(\pt)$ (or $[D(p_T)]^{(3)}$) by the following
  Jacobian factor:
\begin{equation}
\label{eq:jacob}
D(\pt)\to {\cal J}_Q D\left(\pt\right),\qquad {\cal J}_Q \equiv\frac{\left(Q/\pt\right)^p}{1+\left(Q/\pt\right)^p}\,.
\end{equation}
\end{itemize}
On the other hand, one has some freedom of setting the corresponding
scales in the differential NLO cross section for \FJ{} production in
Eq.~\eqref{eq:Bbar}, as long as they tend to $\pt$ at small transverse
momentum. We will discuss possible choices at the end of this section.

In Eq.~\eqref{eq:scales}, $K_{R,F}$ are scale variation parameters
that are varied between $1/2$ and $2$ to estimate perturbative
uncertainties. 
As a way of smoothly approaching non-perturbative scales at low $\pt$,
we introduce the alternative scale setting
\begin{equation}
\label{eq:scales_Q0}
\muR = \KR \,\left(Q \,e^{-L}+Q_0 \,g(\pt)\right)\,,\qquad \muF = \KF
\,\left(Q\, e^{-L} + Q_0 \,g(\pt)\right),
\end{equation}
where $g(\pt)$ is a damping function. The scale $Q_0$ is a
non-perturbative parameter which has the role of regularising the
Landau singularity and as such it should be tuned together with the
hadronization model using experimental data.
One has some freedom in choosing $g(\pt)$, and we explore
the options 
\begin{equation}
\label{eq:gpt}
g(\pt) = 1,\qquad g(\pt) = \frac{1}{1+\frac{Q}{Q_0} e^{-L}}\,.
\end{equation}
The difference between the two is that the second option further
suppresses the shift by $Q_0$ at large transverse momentum.
This prescription is also consistently adopted in the Sudakov form
factor $\tilde{S}(\pt)$, defined in Eq.~\eqref{eq:Rdef}, at the
integrand level. The modified Sudakov is then evaluated exactly via a
numerical calculation of the integral.
As far as the $D(\pt)$ function is concerned, analogously to what has
been discussed for the modified logarithms in Eq.~\eqref{eq:jacob}, the
choice in Eq.~\eqref{eq:scales_Q0} requires the introduction of an additional
factor ${\cal J}_{Q_0}$ (for the two choices in Eq.~\eqref{eq:gpt}, respectively)
\begin{equation}
{\cal J}_{Q_0} \equiv \frac{Q\,e^{-L}}{Q\,e^{-L}+Q_0}\,,\qquad {\cal
  J}_{Q_0} \equiv Q \,e^{-L} \frac{1 -g^2(\pt)}{Q \,e^{-L} +
  Q_0 \, g(\pt)}\,,
\end{equation}
which multiplies only the derivative of the luminosity.\footnote{Since
  the lower bound of integration of the Sudakov form factor is
  unchanged, its derivative simply amounts to
  Eq.~\eqref{eq:Rdef-derivative} with the scale of the coupling set as
  in Eq.~\eqref{eq:scales_Q0}.} This modifies Eq.~\eqref{eq:jacob} as
\begin{equation}
  {\cal J}_Q D(\pt)  \to {\cal J}_Q \left(-\frac{\mathd \tilde{S}(\pt)}{\mathd \pt} {\cal L}(\pt)+{\cal J}_{Q_0}\frac{\mathd {\cal L}(\pt)}{\mathd \pt}\right)\,,
\end{equation}
where the scales are set as in Eq.~\eqref{eq:scales_Q0} after taking
the derivatives in the right-hand-side of the above equation.

The scale $Q_0 > \Lambda$ smoothly freezes the coupling and PDFs at
low scales. We stress that this prescription does not affect the
double logarithmic terms in $\tilde{S}(\pt)$, which ensures
that Eq.~\eqref{eq:Bbar} still vanishes exponentially for $\pt\to 0$. With
this prescription to regularize the Landau pole, we can now integrate
safely all the way down to $\pt$ scales at which the integrand
vanishes, which is an essential requirement in order for
Eq.~\eqref{eq:start} to be NNLO accurate. This is because the
contribution of the total derivative to the integral over $\pt$ must
vanish at the lower bound of integration. If this is not the case,
one introduces an additional systematic uncertainty due to the
truncation (or slicing) of the integral at scales where the integrand
is not vanishingly small. In this respect the scale $Q_0$ is not a
slicing parameter, as it simply acts by freezing the coupling and the
PDFs in the infrared region. At the same time, this allows us to
perform a consistent scale variation all the way down to $\pt= 0$,
which would not be the case if a slicing cutoff were introduced in the
integral of Eq.~\eqref{eq:start}.
We do not find a visible difference between the two options in
Eq.~\eqref{eq:gpt}, and we therefore stick to the second of the two
with $Q_0=2$ GeV as our default.
We stress that, for differential distributions sensitive to infrared
dynamics (for instance the transverse momentum of the $Z$ boson in the
peak region), the $Q_0$ parameter must be determined together with a
tune of the parton-shower hadronization model.

As a final step, we discuss the scale setting adopted in the NLO
\FJ{} cross section in Eq.~\eqref{eq:Bbar}. The default prescription
in \minlo{} and \minnlo{} is to set the perturbative scales in this
term to
\begin{equation}
\muR = \KR \,\pt\,,\qquad \muF = \KF \,\pt\,,
\end{equation}
or, if the smooth freezing is introduced, to
\begin{equation}
\label{eq:scales_orig}
\muR = \KR \,\left(\pt+g(\pt)\,Q_0\right)\,,\qquad \muF = \KF \,\left(\pt+g(\pt)\,Q_0\right)\,.
\end{equation}
This ensures that in the small $\pt$ limit these scales match the ones
used in the Sudakov form factor and the $D(\pt)$ function, which are
constrained by the structure of $\pt$ resummation, hence guaranteeing
a correct matching at small $\pt$. However, one can also choose to set
the scales of the NLO calculation as in
Eqs.~\eqref{eq:scales} and~\eqref{eq:scales_Q0}, such that at large $\pt$
the scales of the \minnlo{} predictions are of the order of the
invariant mass $Q$ of the colour singlet. We employ \eqn{eq:scales_Q0}  in the
results shown in this article. While this choice is more appropriate
for inclusive observables, the one in \eqn{eq:scales_orig} is preferable to obtain
predictions in regimes where the colour singlet is produced with large
$\pt$.
Both options~\eqref{eq:scales_Q0} and~\eqref{eq:scales_orig} are made
available to the user.

\subsection{Impact of shower recoil scheme on kinematics of the colour singlet}
Another source of higher-order corrections in the final
prediction is given by the parton shower. Shower simulations are
expected to be accurate in configurations dominated by soft and/or
collinear radiation.
Away from these limits the corrections introduced by the
shower evolution are subleading (higher order in nature) to the fixed-order description of the hard scattering process.

% \begin{figure}[htpb]
 \begin{figure}[t]
  \centering
  \includegraphics[width=0.6\textwidth]{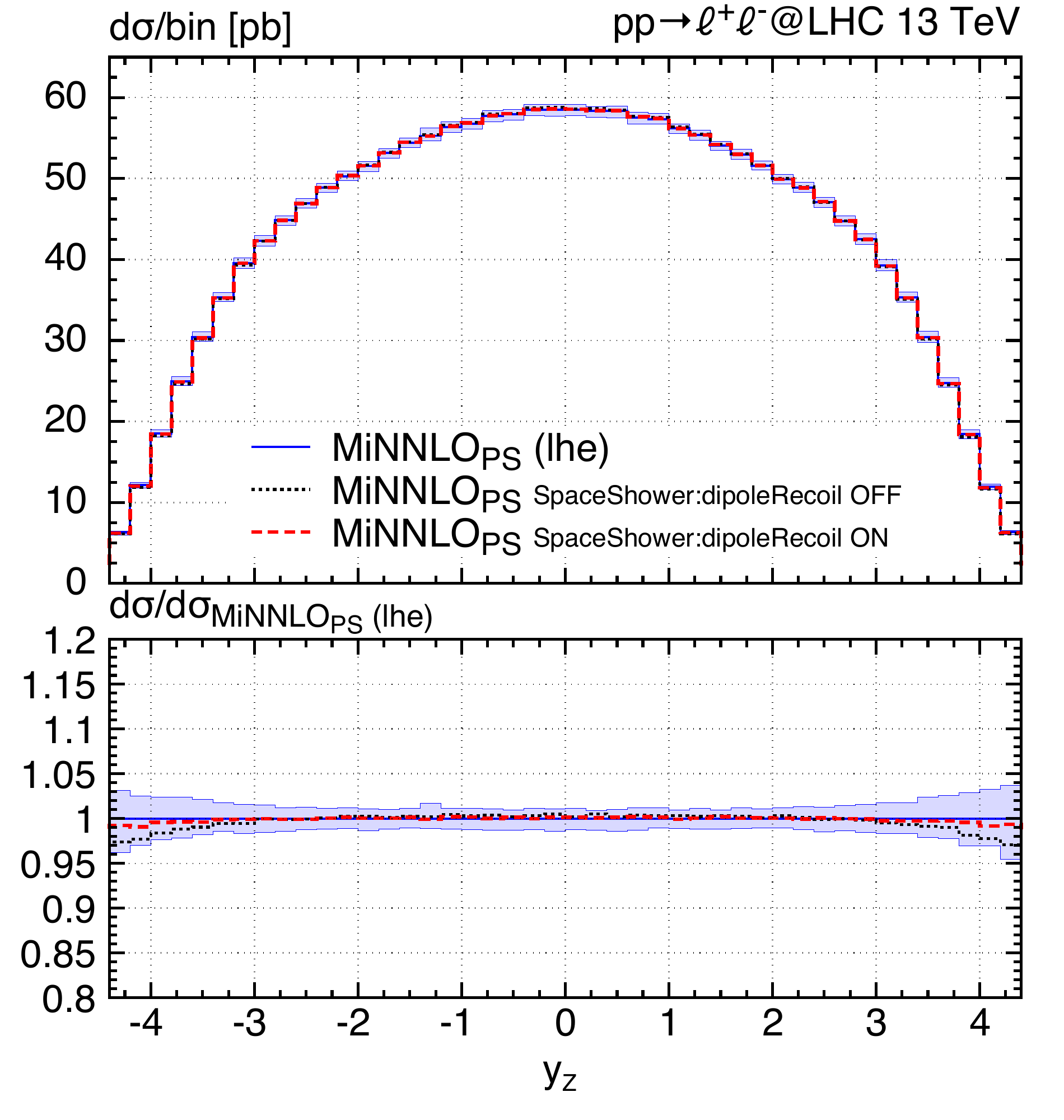}
  \caption{Rapidity distribution of the $Z$ boson.
  The plot compares \minnlo{} predictions at LHE level (blue, solid)
  and after showering using different recoil schemes in \PYTHIA{8} 
    with the option {\tt SpaceShower:dipoleRecoil 0} (black, dotted)
    and {\tt SpaceShower:dipoleRecoil 1} (red, dashed).
    The lower panel shows the ratio to the prediction at LHE level.}
  \label{fig:recoil_comparisons}
\end{figure}

As a consequence, the shower may lead to higher-order corrections to
physical observables that one would naively expect to be largely
insensitive to the showering process. As an example, let us consider
again the rapidity distribution of the $Z$ boson. This inclusive
quantity should be nearly independent of the infrared
dynamics. However, if we compare \minnlo{} predictions for this
quantity at the LHE level and after showering with
\PYTHIA{8}~\cite{Sjostrand:2014zea} (without hadronization) in
\fig{fig:recoil_comparisons}, we observe that the shower suppresses
configurations where the vector boson is produced at large absolute
rapidities.  This effect can traced back to the default (black, dotted
line in Fig.~\ref{fig:recoil_comparisons}) shower recoil scheme used
by \PYTHIA{8} for initial-state radiation, which is global, i.e.\ the
recoil of a generated particle is shared among all particles in the
final state of the event. Naturally, this affects also the kinematics
of the $Z$ boson and it is responsible for the behaviour observed in
Fig.~\ref{fig:recoil_comparisons}.

This observation can be confirmed by choosing an alternative recoil
scheme. In a large-$N_c$ picture, one can for instance share the
recoil globally only for emissions off {\it initial--initial} colour
dipoles, while assigning it locally (i.e.\ entirely taken from a
single final state particle) for emissions off {\it initial--final}
colour dipoles (i.e.\ a colour line that connects an initial-state and
a final-state particle). This scheme is available within \PYTHIA{8}
via the flag {\tt SpaceShower:dipoleRecoil 1} (cf.\
\citere{Cabouat:2017rzi} for details).
In this case, we expect the $Z$ boson to be less affected by the
parton shower than in the default recoil scheme. As an example let us
consider the leading order configuration $q + \bar q \to Z$. If the
quark lines emit a real gluon ($q + \bar q \to Z+g$), the recoil is
taken from the $Z$ boson. One then is left with two initial--final and
no initial--initial colour dipoles. As a consequence, the emission of
an additional radiation will never affect the kinematics of the $Z$
boson. Conversely, an extra radiation from the $q + g \to Z+q$
configuration will affect the $Z$ boson if it is emitted off the
initial-state $qg$ dipole.
From the plot we indeed observe that this {\it less global} version of
the recoil scheme (red, dashed line in
Fig.~\ref{fig:recoil_comparisons}) impacts the $Z$-boson kinematics
very mildly.

We stress that the effects of the shower recoil scheme on the rapidity
distribution are by all means subleading and formally beyond NNLO
accuracy. On the other hand, the choice of the recoil scheme can have
consequences for the logarithmic accuracy of a parton
shower~\cite{Nagy:2009vg,Dasgupta:2018nvj,Bewick:2019rbu}, which
implies that a comprehensive discussion about a given scheme must take
place in this context. 
Specifically, the alternative recoil scheme that we have just
discussed may arguably have consequences for the description of the
transverse-momentum distribution of the $Z$ boson, which in this scheme becomes
insensitive to some of the radiation emitted off the initial-state
quarks. In a transverse-momentum ordered shower like \PYTHIA{8}, this
may result in next-to-leading logarithmic contributions to the $Z$
transverse-momentum spectrum being potentially mistreated
(cf.\ \citere{Dasgupta:2018nvj} for details). Since in this article
we assume parton showers to be LL accurate, this problem is strictly
of subleading nature.
However, recent progress in formulating NLL
accurate parton showers~\cite{Dasgupta:2020fwr,Forshaw:2020wrq} raises
the question of whether the \minnlo{} method (in fact any of the
available \nnlops{}
methods~\cite{Hamilton:2012rf,Alioli:2013hqa,Hoeche:2014aia,Monni:2019whf})
preserves the shower accuracy after matching. A study of this type is
beyond the scope of this article and left for future
work.

As far as the matching to NNLO QCD for the $2\to 1$ processes studied
in this paper is concerned, we use the option {\tt
  SpaceShower:dipoleRecoil 1} as the default for our results so that
shower effects on inclusive quantities are minimised.

\section{Results for Drell Yan and Higgs boson production}
\label{sec:results}

In this section we compare the \nnlops{} predictions obtained with
\minnlo{} to \nnlo{} results obtained with the
public code \Matrix{}~\cite{Grazzini:2017mhc}.
We consider the processes
\begin{align}
pp \rightarrow \ell^+\ell^-\,,\quad pp \rightarrow \ell^-\bar\nu_\ell\,,\quad pp \rightarrow \ell^+\nu_\ell \,, \quad {\rm and} \quad pp \rightarrow H\,,
\end{align}
for massless leptons $\ell\in\{e,\mu\}$. The Higgs is produced on-shell in the 
heavy-top approximation, while for the DY processes the full off-shell effects are
taken into account, including $Z$-boson, $W$-boson, and photon ($\gamma^*$) contributions. For neutral-current DY we restrict the invariant mass of dilepton pair to the $Z$-mass window
\begin{equation}
66~{\rm GeV}\leq M_{\ell^+\ell^-}\leq 116~{\rm GeV}\,
\end{equation}
to avoid the photon singularity.

We consider 13\,TeV LHC collisions. For the EW parameters we employ
the $G_\mu$ scheme with the EW mixing angles given by
$\cos^2\theta_{\tiny{\mbox{W}}}=\mw{}^2/\mz{}^2$ and
$\alpha=\sqrt{2}\,G_\mu \mw{}^2\sin^2\theta_{\tiny{\mbox{W}}}/\pi$.
The following values are used as input parameters:
$G_{\tiny{\mbox{F}}} =1.16639\times 10^{-5}$\,GeV$^{-2}$,
$\mw{}=80.385$\,GeV, $\Gamma_W=2.0854$\,GeV, $m_Z = 91.1876$\,GeV,
$\Gamma_Z=2.4952$\,GeV, and $\mh{} = 125$\,GeV.  With an on-shell
top-quark mass of $\mt= 173.2$\,GeV and $n_f=5$ massless quark
flavours, we use the corresponding NNLO PDF set with $\as(\mz{}) =
0.118$ of NNPDF3.1~\cite{Ball:2017nwa} for the DY results and
the set PDF4LHC15\_nnlo\_mc of PDF4LHC15~\cite{Butterworth:2015oua,Ball:2014uwa,Harland-Lang:2014zoa,Dulat:2015mca}
for Higgs boson production.

The reference \nnlo{} results of \Matrix{} have been obtained by setting the central scales to the invariant mass of the produced color singlet, i.e.\
\begin{equation}
\muR=\muF = Q,\,\qquad Q=M_{\ell^+\ell^-}, M_{\ell^-\bar\nu_\ell}, M_{\ell^+\nu_\ell}, m_H\,,
\label{eq:scales_matrix}
\end{equation}
while the \minnlo{} simulations are obtained using the default setup
discussed in \sct{sec:implementation}.
Scale uncertainties are obtained by varying
the renormalisation and factorisation scales by a factor of two about
their central value while keeping $1/2\leq \muR/\muF \leq 2$.
All \minnlo{} results are showered with \PYTHIA{8}~\cite{Sjostrand:2014zea}, switching off hadronization and underlying event.\footnote{In the codes released with this paper, the \POWHEG{} matching is performed with the option {\tt doublefsr 1} \cite{Nason:2013uba}. This provides a symmetric treatment of the $q\to qg$ and $g\to q\bar{q}$ 
final-state splittings in 
the definition of the starting scale of the shower. This ensures a proper treatment
of observables sensitive to radiation off such configurations. 
We have checked explicitly that the observables considered within this paper 
are unaffected by that option.}
In all of the results that follow, the NNLO prediction of \Matrix{} is
represented by a red, dashed curve with a red band, while the
\minnlo{} prediction is shown in blue, solid.

\subsection{Neutral-current and charged-current Drell Yan production}

{\renewcommand{\arraystretch}{1.5}
\begin{table*}[htp!]
  \vspace*{0.3ex}
  \begin{center}
\begin{small}
%    \begin{tabular}{ll@{\hspace{12mm}}l@{\hspace{12mm}}l}
    \begin{tabular}{||c|c|c|c||}
\hline\hline
Process & NNLO (\Matrix{}) & \minnlo{} & Ratio\\
\hline\hline
$pp \rightarrow \ell^+\ell^- $ & $\;\,1919(1)_{-1.1\%}^{+0.8\%}$\,pb& $\;\,1926(1)_{-1.1\%}^{+1.4\%}$\,pb & 1.004\\
\hline
$pp \rightarrow \ell^-\bar\nu_\ell $ & $\;\,8626(4)_{-1.2\%}^{+1.0\%}$\,pb & $\;\,8689(4)_{-1.5\%}^{+1.7\%}$\,pb & 1.007\\
\hline
$pp \rightarrow \ell^+\nu_\ell $ & $11677(5)_{-1.3\%}^{+0.9\%}$\,pb & $11755(5)_{-1.6\%}^{+1.5\%}$\,pb & 1.007\\
\hline\hline
    \end{tabular}
\end{small}
  \end{center}
  \caption{
    Total cross sections of the Drell Yan production processes. The number in brackets denotes the numerical uncertainty on the last digit.}
\label{tab:DY_xs}
\end{table*}
}

We start by discussing the total production rates of the DY processes, reported in
Table~\ref{tab:DY_xs}. We observe an excellent agreement between the
NNLO QCD prediction and the \minnlo{} result, which are consistent at the
few-permille level. We stress again that the two calculations use different
scale settings and are therefore expected to
differ by effects beyond NNLO. As one can see from
Table~\ref{tab:DY_xs}, these differences are small and the 
central prediction of each calculation lies within the
perturbative uncertainty of the other. Moreover, we observe
that the \minnlo{} calculation features a slightly larger scale
uncertainty. This is due to the more conservative uncertainty
prescription adopted in the \minnlo{} case, which involves varying the
renormalisation scale $\muR$ also in the Sudakov form factor
$\tilde{S(\pt)}$, defined in Eq.~\eqref{eq:Rdef}. This choice better
reflects the perturbative uncertainty associated with the \minnlo{}
matching procedure.

 \begin{figure}
  \centering
  \includegraphics[width=0.48\textwidth]{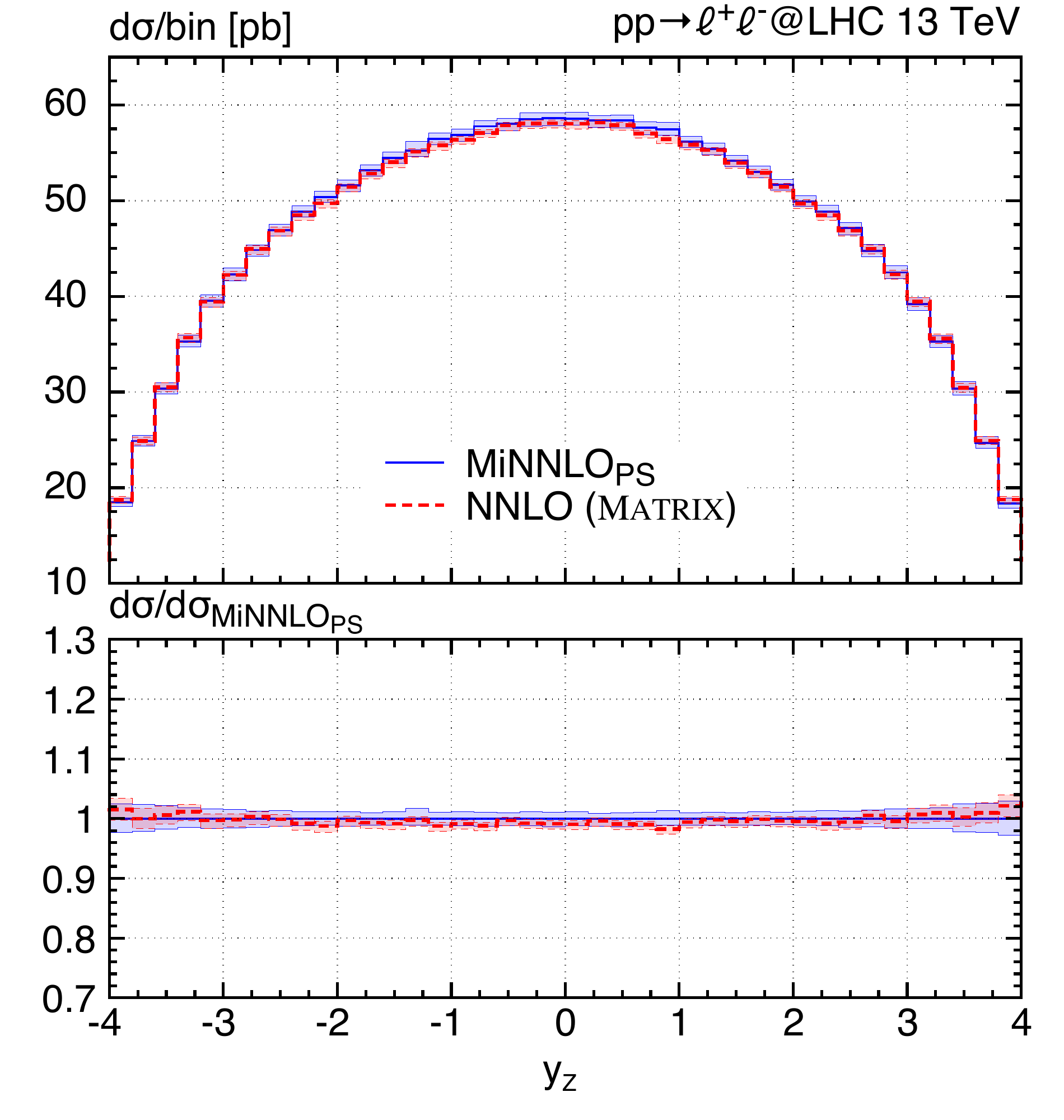}
  \includegraphics[width=0.48\textwidth]{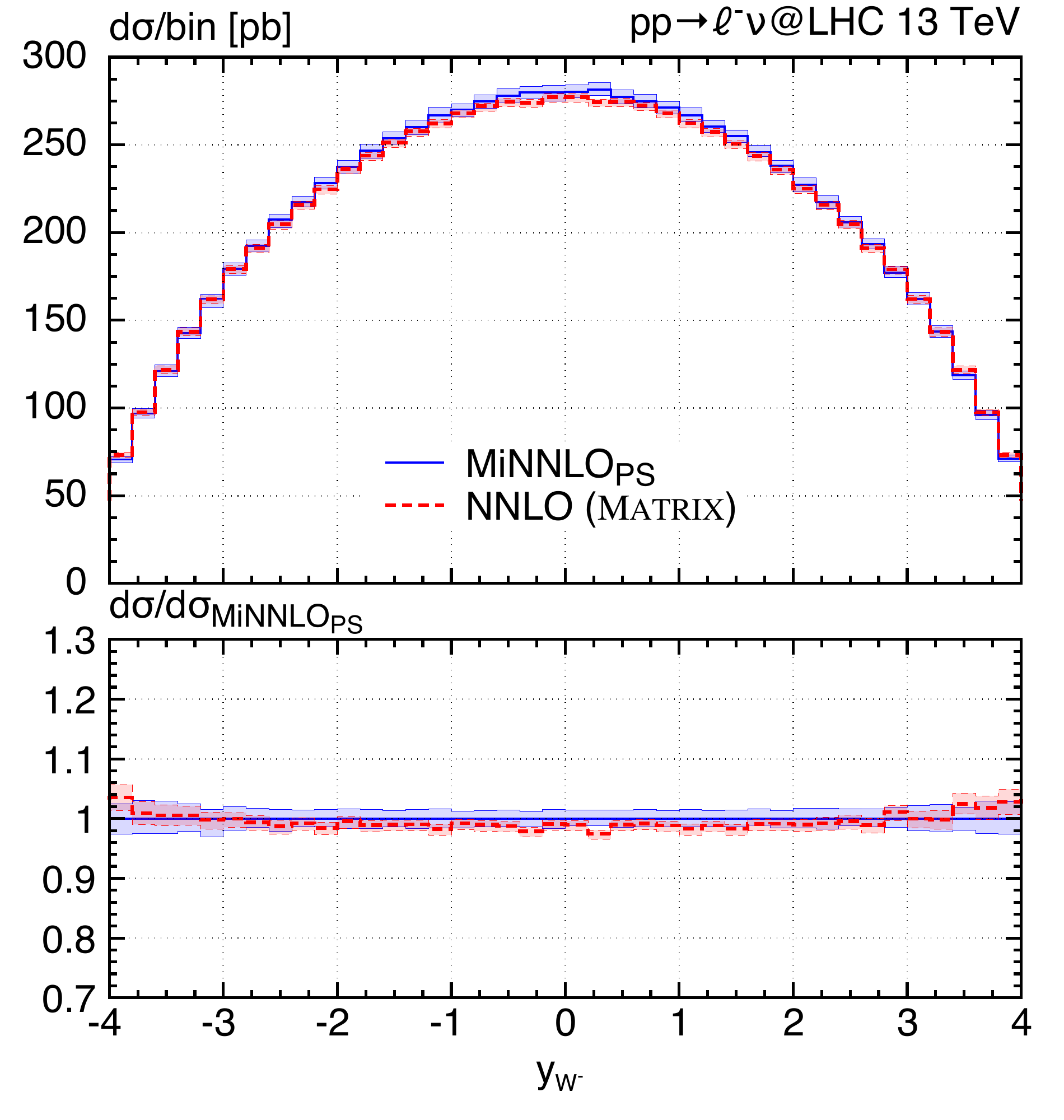}
  \caption{The rapidity distribution of the leptonic pair in neutral-
    (left plot) and charged-current (right plot) Drell Yan production. The lower
    panel shows the ratio of the NNLO and the \minnlo{} predictions
    to the latter.}
  \label{fig:DY_yV}
\end{figure}

\begin{figure}
  \centering
  \includegraphics[width=0.48\textwidth]{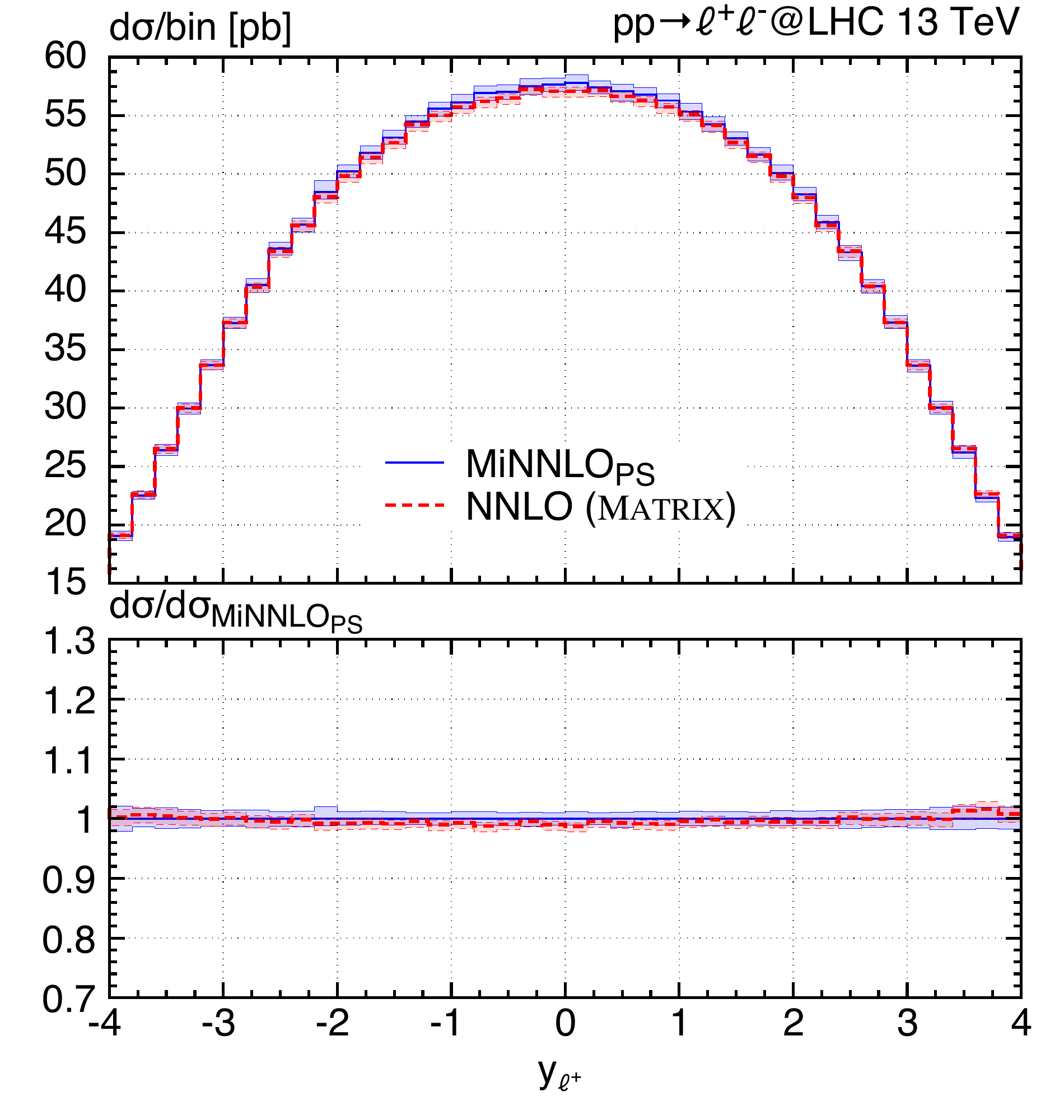}
  \includegraphics[width=0.48\textwidth]{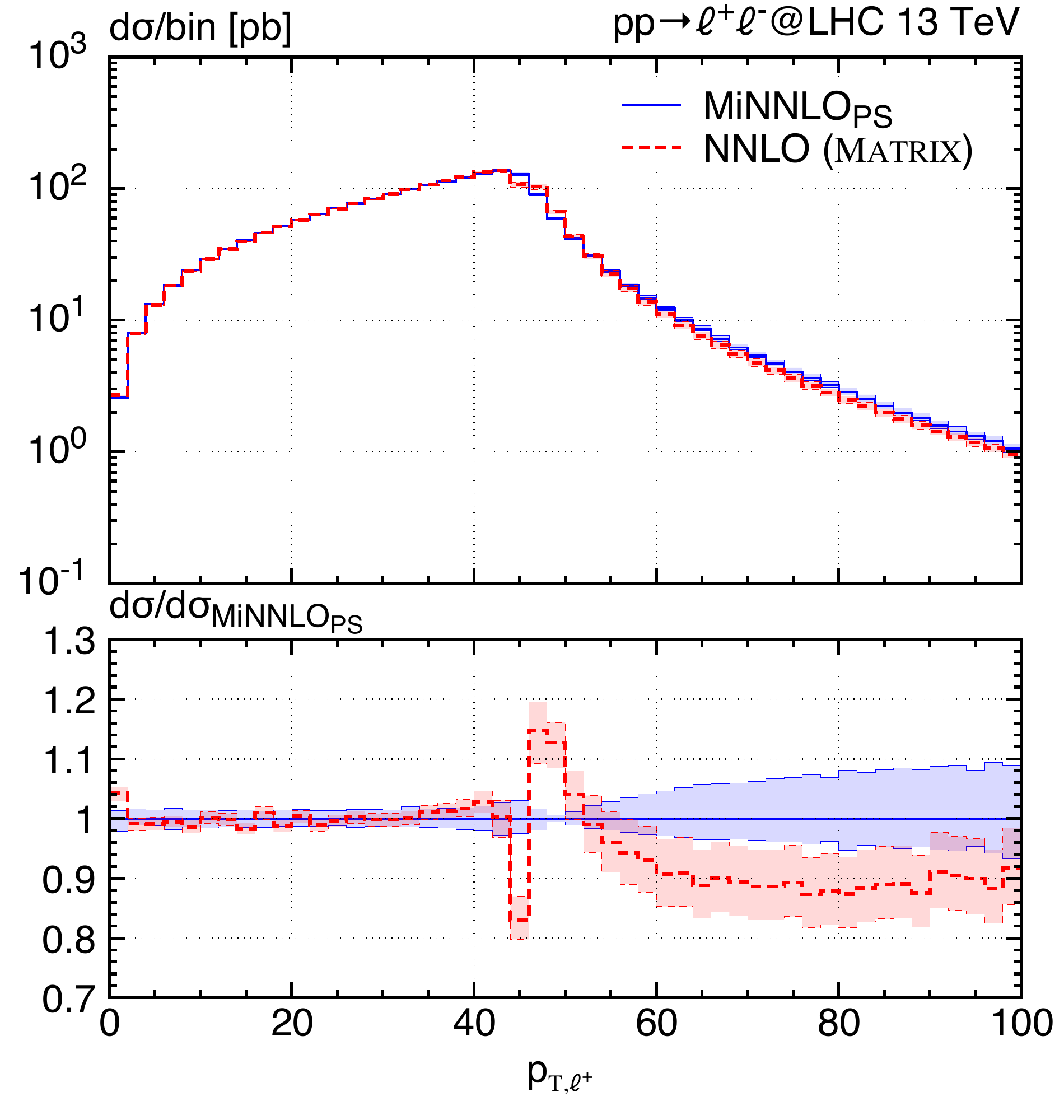}
  \caption{Rapidity distribution (left) and transverse momentum
    (right) of the positively charged lepton in neutral-current Drell
    Yan production. The lower panel shows the ratio to the \minnlo{}
    prediction.}
  \label{fig:DY_Zlep}
\end{figure}
%\vspace{1cm}
 \begin{figure}[t]
  \centering
  \includegraphics[width=0.48\textwidth]{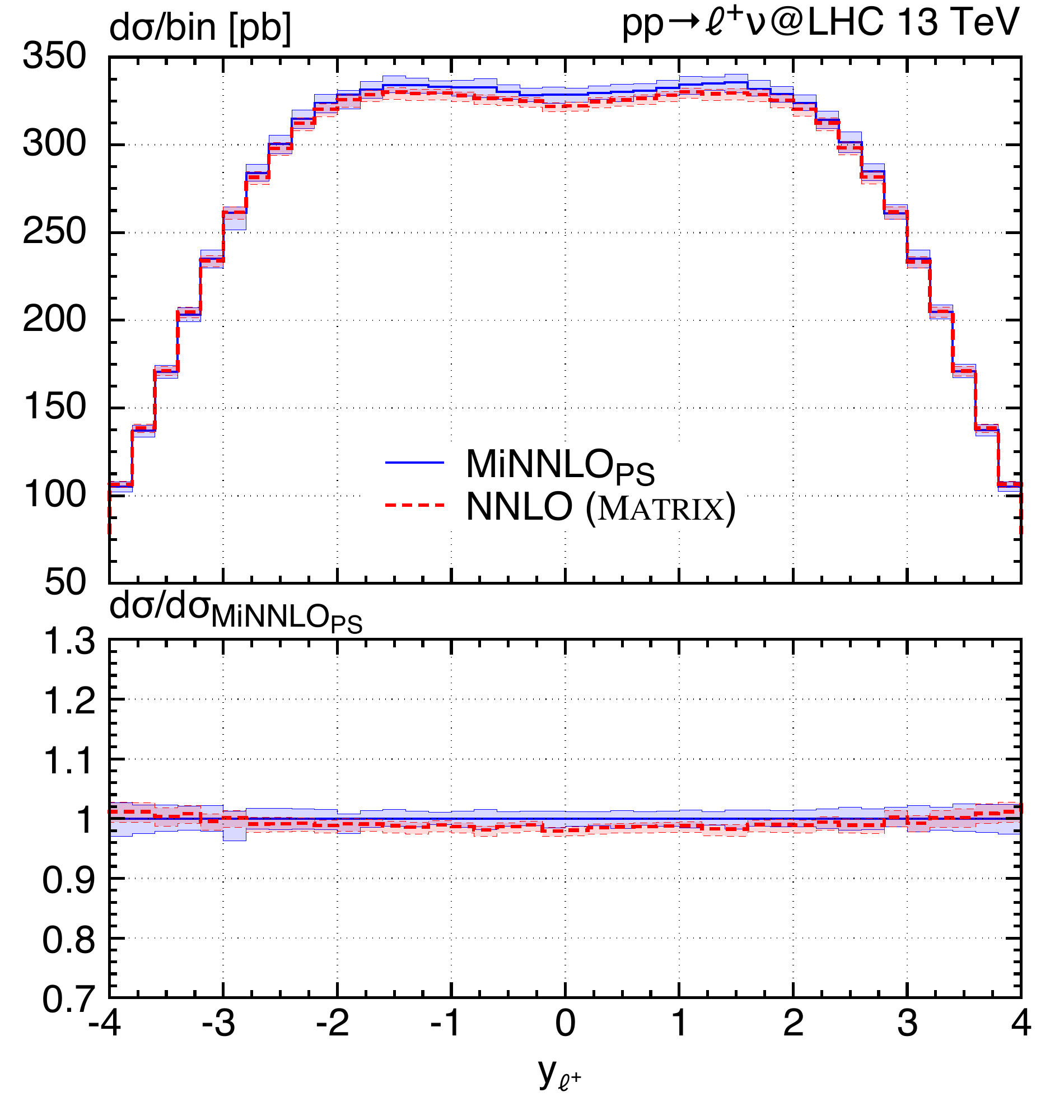}
  \includegraphics[width=0.48\textwidth]{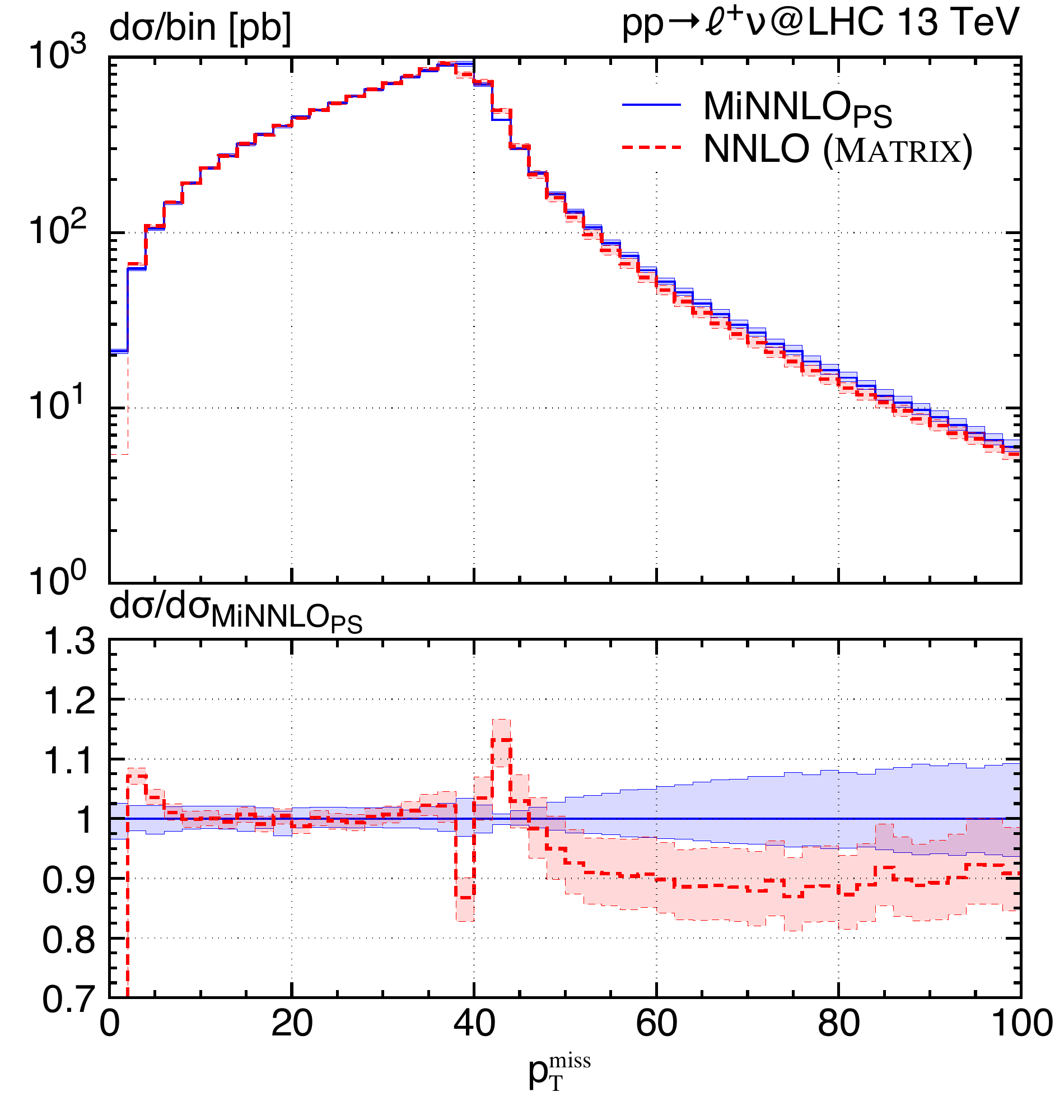}
  \caption{Rapidity distribution of the charged lepton (left) and
    missing transverse momentum (right) in charged-current Drell Yan
    production. The lower panel shows the ratio to the \minnlo{}
    prediction.}
  \label{fig:DY_Wlep}
\end{figure}

We continue by considering the rapidity distribution of the leptonic
system in $Z$/$\gamma^*$ and $W^-$ production, shown in
Fig.~\ref{fig:DY_yV}.  The considerations made above for the inclusive
cross section hold in this case as well, and we observe a very good
agreement between the \minnlo{} and the \nnlo{} predictions across the
entire spectrum, with moderately larger perturbative uncertainties in
the \minnlo{} case.
In comparison to the $Z$ rapidity distribution presented in
Ref.~\cite{Monni:2019whf}, we observe that the shape of the new
\minnlo{} result is much closer to the \nnlo{} prediction in the
forward rapidity region.
Each of the aspects discussed in this article (reduced difference due
to higher-order terms with respect to \nnlo{}, improved evolution of the PDFs
and scale setting, choice of the shower recoil scheme) plays a role in
this improvement, as discussed in the previous section.

Finally, we show a sample of kinematic distribution of the
final-state leptons.  For neutral-current DY production
we compare \minnlo{} to \nnlo{} predictions for the rapidity distribution and
the transverse-momentum distribution of the positively charged lepton
in Fig.~\ref{fig:DY_Zlep}. Similarly, in the case of $W^+$ production
we show the same comparison for the missing transverse-momentum
distribution and for the rapidity distribution of the charged lepton
in Fig~\ref{fig:DY_Wlep}. We observe a very good agreement between the
two calculations for the rapidity distributions, and for the region of
the transverse-momentum spectrum insensitive to shower effects.
Conversely, the parton shower provides an improved description for
$p_{\rm T,\ell^+}~(p_{\rm T}^{\rm miss}) \lesssim 5$\,GeV and
$p_{\rm T,\ell^+}~(p_{\rm T}^{\rm miss}) \gtrsim m_V/2$
%, with $V\in\{Z,W^+\}$ being the respective vector boson, 
where the cross
section is sensitive to multi particle emissions and therefore
receives relevant corrections from the parton shower that resums
integrable, but large logarithmic terms.  The perturbative instability
at the threshold is a well known feature of fixed-order
calculations~\cite{Catani:1997xc}. It appears at $p_{\rm T,\ell^+}~(p_{\rm T}^{\rm miss}) \sim m_V/2$,
since at LO, where the leptons are back-to-back and can share only the
available partonic centre-of-mass energy $\sqrt{\hat s} = Q$, the
distribution is kinematically restricted to the region $p_{\rm T,\ell^+}~(p_{\rm T}^{\rm miss})\le Q/2$
and on-shell configurations $Q \sim m_V$ provide by far the dominant
contribution.  The region $p_{\rm T,\ell^+}~(p_{\rm T}^{\rm miss}) \gtrsim m_V/2$ is filled only upon
inclusion of higher-order corrections, and the NNLO predictions
becomes effectively only NLO accurate, as indicated by the enlarged
uncertainty bands.

\newpage
\subsection{Higgs boson production}

{\renewcommand{\arraystretch}{1.5}
\begin{table*}[htp!]
  \vspace*{0.3ex}
  \begin{center}
\begin{small}
%    \begin{tabular}{ll@{\hspace{12mm}}l@{\hspace{12mm}}l}
    \begin{tabular}{||c|c|c|c||}
\hline\hline
Process & NNLO (\Matrix{}) & \minnlo{} & ratio \\
\hline\hline
$pp \rightarrow H $ & $39.64(1)_{-10.4\%}^{+10.7\%}$\,pb & $38.03(2)_{-9.0\%}^{+10.2\%}$\,pb & 0.960 \\
\hline\hline
    \end{tabular}
\end{small}
  \end{center}
  \caption{Total cross sections of Higgs-boson production. The number in brackets denotes the numerical uncertainty on the last digit.}
\label{tab:H_xs}
\end{table*}
}

Table~\ref{tab:H_xs} gives the inclusive Higgs cross section at \nnlo{}
computed with \Matrix{} and the one obtained with the \minnlo{}
generator. As in the case of DY production, we observe a good
agreement between the two predictions that are well compatible within
the quoted scale uncertainties, and they are closer than in the
original setup of \citere{Monni:2019whf}.  The moderate numerical
difference between the two results is due to the different scale
settings in the two calculations.

\begin{figure}[t]
  \centering
  \includegraphics[width=0.48\textwidth]{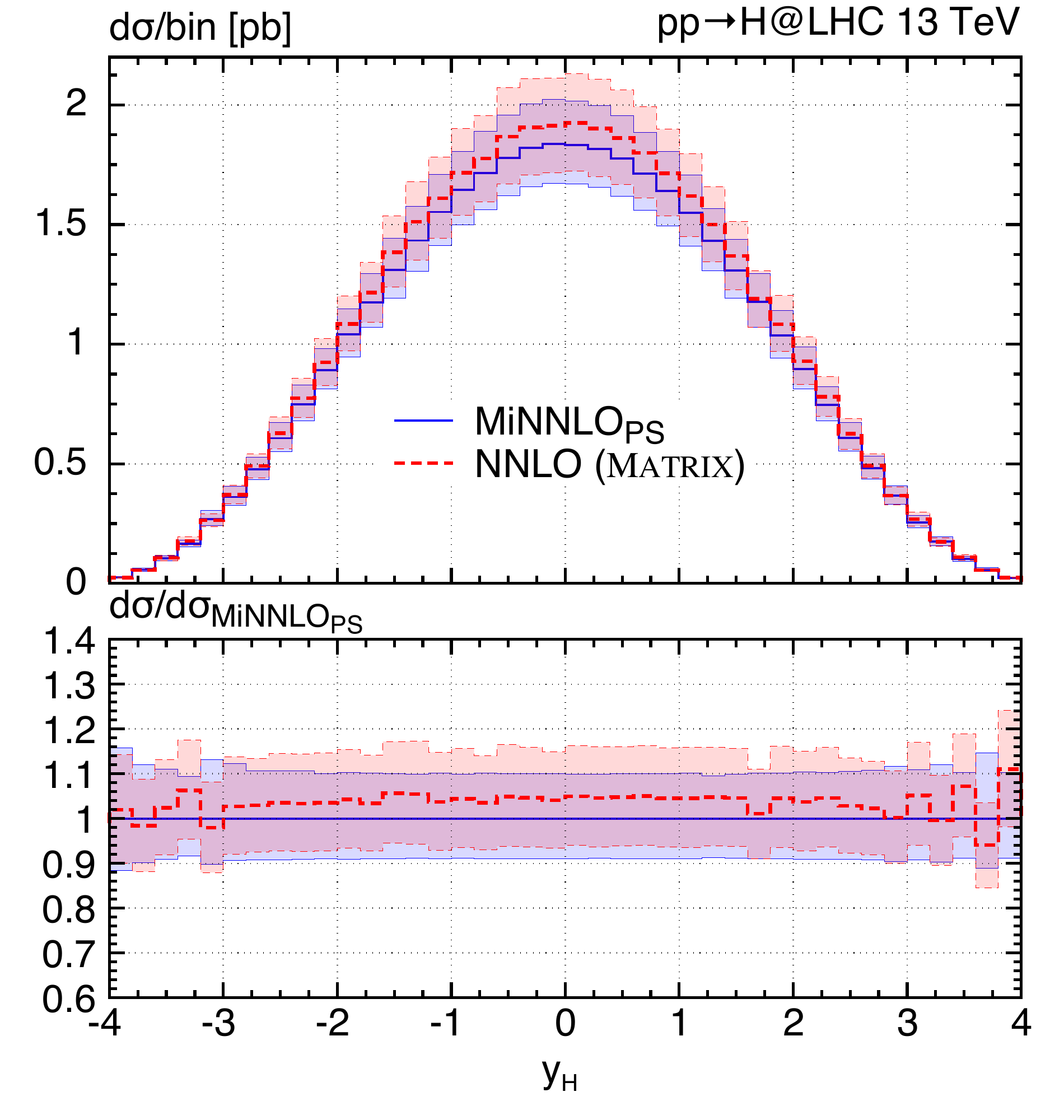}
  \includegraphics[width=0.48\textwidth]{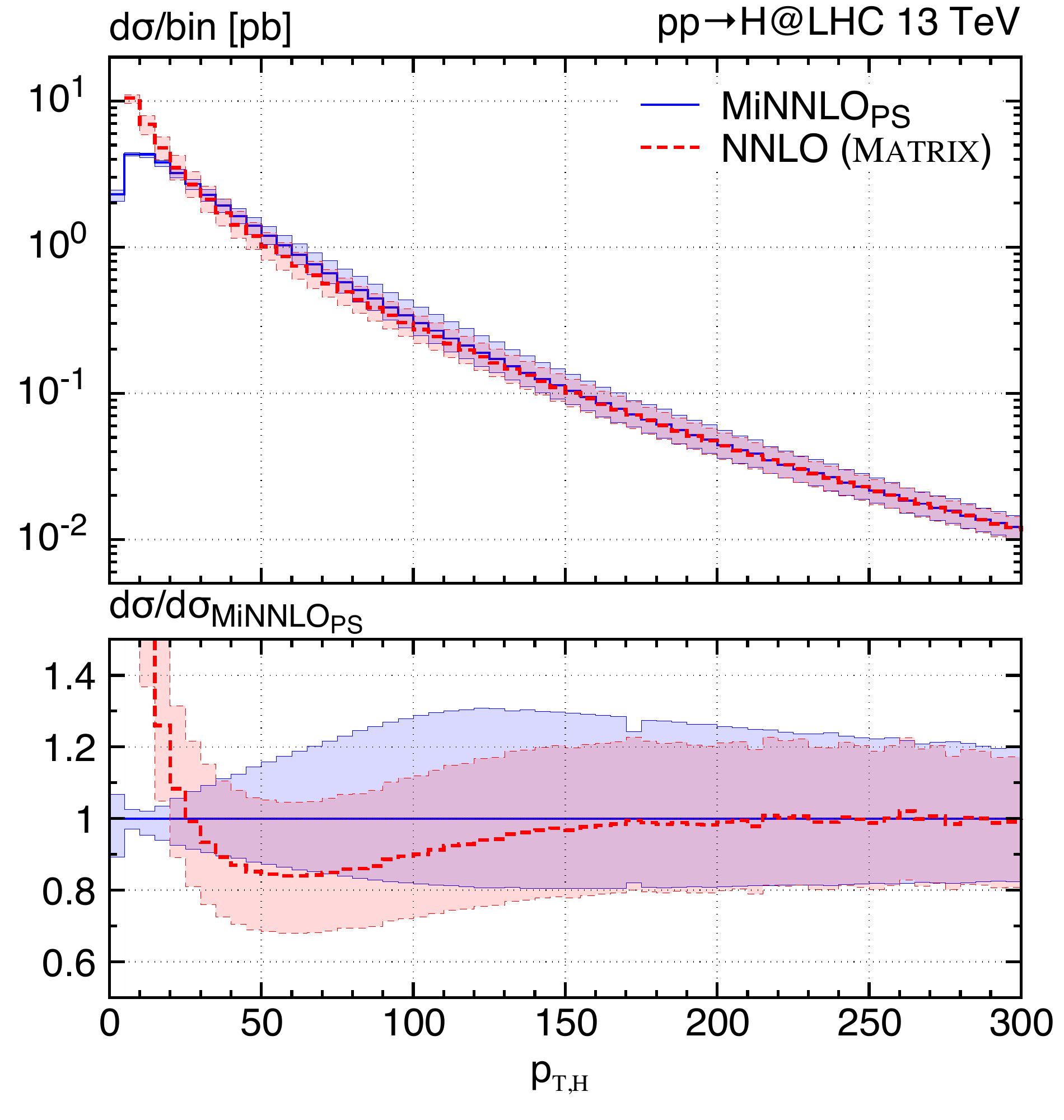}
  \caption{The rapidity distribution of the Higgs boson (left) and its
    transverse momentum (right). The lower panel shows the ratio of
    the NNLO and the \minnlo{} predictions to the latter.}
\label{fig:H_distr}
\end{figure}

The rapidity distribution of the Higgs boson is shown in the left plot
of Fig.~\ref{fig:H_distr}. The \minnlo{} and NNLO
predictions are in mutually good agreement within the perturbative
uncertainties.
The right plot of Fig.~\ref{fig:H_distr} shows the Higgs
transverse-momentum distribution. This observable displays the effect
of the \minnlo{} scale setting in \eqn{eq:scales_Q0} compared to the
one in the \Matrix{} computation in \eqn{eq:scales_matrix}.  The two
scales differ significantly at low and moderate transverse momenta,
while they become identical at large transverse momentum
$p_{\rm T,H}\gtrsim m_H$, where the \minnlo{} and \Matrix{}
predictions are in full agreement.
We recall that the scales of the differential NLO cross section for
\FJ{} production in Eq.~\eqref{eq:Bbar} can also be set to the
transverse momentum as in Eq.~\eqref{eq:scales_orig}. This choice,
used in the original publication~\cite{Monni:2019whf}, is more
appropriate in regimes where the Higgs boson (or the accompanying QCD
jets) are produced with large transverse momentum.

\section{Conclusions}
\label{sec:conclusions}
In this article we have addressed a number of aspects of the \minnlo{}
method, which combines NNLO QCD calculations with parton-shower
simulations. As a case study we have considered the production of a
colour-singlet final state in $2\to 1$ reactions at the LHC.

We have identified the main sources of differences between the \minnlo{}
prediction and \nnlo{} calculations, which are due to
corrections beyond accuracy introduced in the matching procedure.
A number of prescriptions has been presented to either remove or reduce
the impact of such corrections in the \minnlo{} results, specifically:
\begin{itemize}
\item The \minnlo{} formula has been refined to include additional
  terms at all orders in the matching procedure that reduce the
  subleading differences between \minnlo{} and \nnlo{} calculations.
\item The evolution of the parton densities at small transverse
  momentum has been improved, and the scale setting in the coupling
  and PDFs has been consistently adjusted so that at large transverse momentum it
  matches that of the \nnlo{} calculation.
\item We studied the impact of the parton-shower recoil
  scheme on the kinematics of the colour singlet, and discussed how
  this dependence can be reduced.
\end{itemize}
The new prescriptions for the \minnlo{} matching procedure have 
been used to obtain updated predictions for
Higgs, charged- and neutral-current Drell Yan production, finding
a significantly improved agreement between the \minnlo{} and \nnlo{}
calculations for inclusive observables, with commensurate scale
uncertainties.

The prescriptions presented in this article do not affect the
performance of the \minnlo{} method in terms of efficiency and
speed. The generation of fully exclusive \nnlops{} events is merely
$50\%$ slower than the corresponding \minlo{} calculation.
The codes to obtain the results presented in this article are
released within the \POWHEGBOX{} framework.\footnote{The \minnlo{}
  codes can be obtained through downloading the latest revision of the
  {\tt HJ}, {\tt Zj} and {\tt Wj} processes within \POWHEGBOXVtwo{}
  available from \url{http://powhegbox.mib.infn.it}. They are compiled
  directly from the respective {\tt HJMiNNLO}, {\tt ZjMiNNLO} and {\tt
    WjMiNNLO} subfolders of those processes.}
    
\noindent {\bf Acknowledgements.}
We would like to thank Paolo Nason and Giulia Zanderighi for useful
discussions and constructive comments on the manuscript. We also wish
to thank Luca Rottoli for providing us with a toy set of parton
distributions with a low cutoff that we used to validate our PDF
evolution at low factorisation scales.

\appendix
\section{Explicit formulae for the evaluation of $D(\pt)$}
\label{app:formulae}
In this section we supplement the formulae given in
Ref.~\cite{Monni:2019whf} with the ones required to calculate the {\it
  untruncated} variant of the function $[D(\pt)]^{(3)}$ given in
Eq.~\eqref{eq:Dterms_new}.
The luminosity factor is defined as
as~\cite{Monni:2019whf}
\begin{align} 
{\cal L}(\pt)&=\sum_{c,
c'}\frac{\mathd|M^{\scriptscriptstyle\rm F}|_{cc'}^2}{\mathd\PhiB} \sum_{i, j}
\bigg\{\left(\tilde{C}^{[a]}_{c i}\otimes f_i^{[a]}\right) \tilde{H}(\pt)
\left(\tilde{C}^{[b]}_{c' j}\otimes f_j^{[b]}\right) +
\left(G^{[a]}_{c i}\otimes f_i^{[a]}\right) \tilde{H}(\pt) \left(G^{[b]}_{c'
j}\otimes f_j^{[b]}\right)\bigg\}\,.
\label{eq:luminosity}
\end{align}
Here $M^{\scriptscriptstyle\rm F}$ is the Born matrix element for the
production of the colour singlet \F{}, and $f$ denotes the parton
distribution functions. Moreover, $\tilde{H}$ encodes the virtual
corrections to this process up to two loops, and $\tilde{C}$ and $G$
are the coefficient functions up to ${\cal O}(\as^2)$ (see
Ref.~\cite{Monni:2019whf} for details).
The Sudakov form factor $\tilde{S}(\pt)$ is defined as
\begin{equation}
\label{eq:Rdef}
\tilde{S}(\pt) = 2\int_{\pt}^{Q}\frac{\mathd q}{q}
                    \left(A(\as(q))\ln\frac{Q^2}{q^2} +
                    \tilde{B}(\as(q))\right),
\end{equation}
with
\begin{align}
A(\as)=& \left(\abar\right) A^{(1)} + \left(\abar\right)^2 A^{(2)}+ \left(\abar\right)^3 A^{(3)}\,,\notag\\
\tilde{B}(\as)=& \left(\abar\right) B^{(1)} + \left(\abar\right)^2 \tilde{B}^{(2)}\,,
\end{align}
and its derivative reads
\begin{equation}
\label{eq:Rdef-derivative}
\frac{\mathd \tilde{S}(\pt)}{\mathd \pt} = -\frac{2}{\pt}
                    \left(A(\as(\pt))\ln\frac{Q^2}{\pt^2} +
                    \tilde{B}(\as(\pt))\right)\,.
\end{equation}
All coefficients of the above equations are defined in Section 4 and
Appendix B of Ref.~\cite{Monni:2019whf}, including their scale
dependence. The above formulae are inserted into
Eq.~\eqref{eq:Dterms_new}, where the derivative of ${\cal L}(\pt)$ is
evaluated numerically to ensure that the exact total derivative in
\eqn{eq:start} that we started our derivation from is not
modified. The numerical derivative of ${\cal L}(\pt)$ is performed by
evaluating \eqn{eq:luminosity} using a five-point stencil discrete
derivative calculated on a fine grid around $\pt$.

One last necessary ingredient is given by the first and second order
expansion of $D(\pt)$ in powers of $\as(\pt)$. Its coefficients read
\begin{align}
  [D(\pt)]^{(1)} &= 
  -\left[\frac{\mathd \tilde{S}(\pt)}{\mathd \pt}\right]^{(1)}[{\cal L}(\pt)]^{(0)}
+ \left[\frac{\mathd {\cal L}(\pt)}{\mathd \pt}\right]^{(1)},\nonumber\\
  [D(\pt)]^{(2)} &= 
  -\left[\frac{\mathd \tilde{S}(\pt)}{\mathd \pt}\right]^{(2)}[{\cal L}(\pt)]^{(0)}
  -\left[\frac{\mathd \tilde{S}(\pt)}{\mathd \pt}\right]^{(1)}[{\cal
    L}(\pt)]^{(1)}+ \left[\frac{\mathd {\cal L}(\pt)}{\mathd
                   \pt}\right]^{(2)} ,
\end{align}
where
\begin{align}
\left[\frac{\mathd \tilde{S}(\pt)}{\mathd \pt}\right]^{(1)}  &= -\frac2{\pt} \left(A^{(1)}\ln\frac{Q^2}{\pt^2} +
                    B^{(1)}\right)\,,\notag\\
\left[\frac{\mathd \tilde{S}(\pt)}{\mathd \pt}\right]^{(2)}  &= -\frac2{\pt} \left(A^{(2)}\ln\frac{Q^2}{\pt^2} +
                    \tilde{B}^{(2)}\right)\,,\notag\\
[{\cal L}(\pt)]^{(0)} & = \sum_{c,
c'}\frac{\mathd|M^{\scriptscriptstyle\rm F}|_{cc'}^2}{\mathd\Phi_{\rm
                        B}}\,f_c^{[a]}f_{c'}^{[b]}\,,\notag\\
[{\cal L}(\pt)]^{(1)} & = \sum_{c,
c'}\frac{\mathd|M^{\scriptscriptstyle\rm F}|_{cc'}^2}{\mathd\Phi_{\rm
                        B}}\bigg\{H^{(1)}f_c^{[a]}f_{c'}^{[b]} +
                        (C^{(1)}\otimes f)_c^{[a]}f_{c'}^{[b]} + 
                        f_{c}^{[a]} (C^{(1)}\otimes
                        f)_{c'}^{[b]}\bigg\}\,,\notag\\
\left[  \frac{\mathd {\cal L}(\pt)}{\mathd \pt}
  \right]^{(1)} & = \sum_{c,
c'}\frac{\mathd|M^{\scriptscriptstyle\rm F}|_{cc'}^2}{\mathd\Phi_{\rm
                        B}}\frac{2}{\pt}\bigg\{ (\hat{P}^{(0)}\otimes
                             f)_c^{[a]}f_{c'}^{[b]} +  f_{c}^{[a]}
                             (\hat{P}^{(0)}\otimes f)_{c'}^{[b]}\bigg\}\,,\notag\\
\left[  \frac{\mathd {\cal L}(\pt)}{\mathd \pt}
  \right]^{(2)} & = \sum_{c,
c'}\frac{\mathd|M^{\scriptscriptstyle\rm F}|_{cc'}^2}{\mathd\Phi_{\rm
                        B}}\frac{2}{\pt}\bigg\{ (\hat{P}^{(1)}\otimes
                             f)_c^{[a]}f_{c'}^{[b]} +  f_{c}^{[a]}
                             (\hat{P}^{(1)}\otimes
                    f)_{c'}^{[b]}\notag\\
&+H^{(1)}\left[(\hat{P}^{(0)}\otimes
                             f)_c^{[a]}f_{c'}^{[b]} +  f_{c}^{[a]}
                             (\hat{P}^{(0)}\otimes f)_{c'}^{[b]}
                             \right]\notag\\
&+  (C^{(1)}\otimes f)_c^{[a]}(\hat{P}^{(0)}\otimes f)_{c'}^{[b]} +
  (\hat{P}^{(0)}\otimes f)_c^{[a]}(C^{(1)}\otimes
  f)_{c'}^{[b]}\notag\\
&+  f_c^{[a]}(\hat{P}^{(0)}\otimes C^{(1)}\otimes f)_{c'}^{[b]} +
  (\hat{P}^{(0)}\otimes C^{(1)}\otimes f)_c^{[a]}f_{c'}^{[b]}\notag\\
&- 2\beta_0 \pi \Big[H^{(1)}f_c^{[a]}f_{c'}^{[b]} +
                        (C^{(1)}\otimes f)_c^{[a]}f_{c'}^{[b]} + 
                        f_{c}^{[a]} (C^{(1)}\otimes
  f)_{c'}^{[b]}\Big]\bigg\}\,.
\end{align}
The scale dependence is implemented as in Appendix D of
Ref.~\cite{Monni:2019whf}, and in addition the above $ [D(\pt)]^{(1)}
$ and $ [D(\pt)]^{(2)} $ terms depend on the renormalisation and
factorisation scale factors $\KR$ and $\KF$ as
\begin{align}
[D(\pt)]^{(1)}(\KF,\KR) &= [D(\pt)]^{(1)}\,,\notag\\
[D(\pt)]^{(2)}(\KF,\KR) &= [D(\pt)]^{(2)} - 2\beta_0 \pi \left[  \frac{\mathd {\cal L}(\pt)}{\mathd \pt}
  \right]^{(1)} \ln\frac{\KF^2}{\KR^2}\,.
\end{align}
\vspace{0.1cm}
\setlength{\bibsep}{3.5pt}
\renewcommand{\em}{}
\bibliographystyle{apsrev4-1}
\bibliography{MiNNLO}

\begin{thebibliography}{25}%
\makeatletter
\providecommand \@ifxundefined [1]{%
 \@ifx{#1\undefined}
}%
\providecommand \@ifnum [1]{%
 \ifnum #1\expandafter \@firstoftwo
 \else \expandafter \@secondoftwo
 \fi
}%
\providecommand \@ifx [1]{%
 \ifx #1\expandafter \@firstoftwo
 \else \expandafter \@secondoftwo
 \fi
}%
\providecommand \natexlab [1]{#1}%
\providecommand \enquote  [1]{``#1''}%
\providecommand \bibnamefont  [1]{#1}%
\providecommand \bibfnamefont [1]{#1}%
\providecommand \citenamefont [1]{#1}%
\providecommand \href@noop [0]{\@secondoftwo}%
\providecommand \href [0]{\begingroup \@sanitize@url \@href}%
\providecommand \@href[1]{\@@startlink{#1}\@@href}%
\providecommand \@@href[1]{\endgroup#1\@@endlink}%
\providecommand \@sanitize@url [0]{\catcode `\\12\catcode `\$12\catcode
  `\&12\catcode `\#12\catcode `\^12\catcode `\_12\catcode `\%12\relax}%
\providecommand \@@startlink[1]{}%
\providecommand \@@endlink[0]{}%
\providecommand \url  [0]{\begingroup\@sanitize@url \@url }%
\providecommand \@url [1]{\endgroup\@href {#1}{\urlprefix }}%
\providecommand \urlprefix  [0]{URL }%
\providecommand \Eprint [0]{\href }%
\providecommand \doibase [0]{http://dx.doi.org/}%
\providecommand \selectlanguage [0]{\@gobble}%
\providecommand \bibinfo  [0]{\@secondoftwo}%
\providecommand \bibfield  [0]{\@secondoftwo}%
\providecommand \translation [1]{[#1]}%
\providecommand \BibitemOpen [0]{}%
\providecommand \bibitemStop [0]{}%
\providecommand \bibitemNoStop [0]{.\EOS\space}%
\providecommand \EOS [0]{\spacefactor3000\relax}%
\providecommand \BibitemShut  [1]{\csname bibitem#1\endcsname}%
\let\auto@bib@innerbib\@empty
%</preamble>
\bibitem [{\citenamefont {Hamilton}\ \emph {et~al.}(2013)\citenamefont
  {Hamilton}, \citenamefont {Nason}, \citenamefont {Oleari},\ and\
  \citenamefont {Zanderighi}}]{Hamilton:2012rf}%
  \BibitemOpen
  \bibfield  {author} {\bibinfo {author} {\bibfnamefont {K.}~\bibnamefont
  {Hamilton}}, \bibinfo {author} {\bibfnamefont {P.}~\bibnamefont {Nason}},
  \bibinfo {author} {\bibfnamefont {C.}~\bibnamefont {Oleari}}, \ and\ \bibinfo
  {author} {\bibfnamefont {G.}~\bibnamefont {Zanderighi}},\ }\href {\doibase
  10.1007/JHEP05(2013)082} {\bibfield  {journal} {\bibinfo  {journal} {JHEP}\
  }\textbf {\bibinfo {volume} {05}},\ \bibinfo {pages} {082} (\bibinfo {year}
  {2013})},\ \Eprint {http://arxiv.org/abs/1212.4504} {arXiv:1212.4504
  [hep-ph]} \BibitemShut {NoStop}%
%%CITATION = ARXIV:1212.4504;%%
\bibitem [{\citenamefont {Alioli}\ \emph {et~al.}(2014)\citenamefont {Alioli},
  \citenamefont {Bauer}, \citenamefont {Berggren}, \citenamefont {Tackmann},
  \citenamefont {Walsh},\ and\ \citenamefont {Zuberi}}]{Alioli:2013hqa}%
  \BibitemOpen
  \bibfield  {author} {\bibinfo {author} {\bibfnamefont {S.}~\bibnamefont
  {Alioli}}, \bibinfo {author} {\bibfnamefont {C.~W.}\ \bibnamefont {Bauer}},
  \bibinfo {author} {\bibfnamefont {C.}~\bibnamefont {Berggren}}, \bibinfo
  {author} {\bibfnamefont {F.~J.}\ \bibnamefont {Tackmann}}, \bibinfo {author}
  {\bibfnamefont {J.~R.}\ \bibnamefont {Walsh}}, \ and\ \bibinfo {author}
  {\bibfnamefont {S.}~\bibnamefont {Zuberi}},\ }\href {\doibase
  10.1007/JHEP06(2014)089} {\bibfield  {journal} {\bibinfo  {journal} {JHEP}\
  }\textbf {\bibinfo {volume} {06}},\ \bibinfo {pages} {089} (\bibinfo {year}
  {2014})},\ \Eprint {http://arxiv.org/abs/1311.0286} {arXiv:1311.0286
  [hep-ph]} \BibitemShut {NoStop}%
%%CITATION = ARXIV:1311.0286;%%
\bibitem [{\citenamefont {Höche}\ \emph {et~al.}(2015)\citenamefont {Höche},
  \citenamefont {Li},\ and\ \citenamefont {Prestel}}]{Hoeche:2014aia}%
  \BibitemOpen
  \bibfield  {author} {\bibinfo {author} {\bibfnamefont {S.}~\bibnamefont
  {Höche}}, \bibinfo {author} {\bibfnamefont {Y.}~\bibnamefont {Li}}, \ and\
  \bibinfo {author} {\bibfnamefont {S.}~\bibnamefont {Prestel}},\ }\href
  {\doibase 10.1103/PhysRevD.91.074015} {\bibfield  {journal} {\bibinfo
  {journal} {Phys. Rev.}\ }\textbf {\bibinfo {volume} {D91}},\ \bibinfo {pages}
  {074015} (\bibinfo {year} {2015})},\ \Eprint {http://arxiv.org/abs/1405.3607}
  {arXiv:1405.3607 [hep-ph]} \BibitemShut {NoStop}%
%%CITATION = ARXIV:1405.3607;%%
\bibitem [{\citenamefont {Monni}\ \emph {et~al.}(2019)\citenamefont {Monni},
  \citenamefont {Nason}, \citenamefont {Re}, \citenamefont {Wiesemann},\ and\
  \citenamefont {Zanderighi}}]{Monni:2019whf}%
  \BibitemOpen
  \bibfield  {author} {\bibinfo {author} {\bibfnamefont {P.~F.}\ \bibnamefont
  {Monni}}, \bibinfo {author} {\bibfnamefont {P.}~\bibnamefont {Nason}},
  \bibinfo {author} {\bibfnamefont {E.}~\bibnamefont {Re}}, \bibinfo {author}
  {\bibfnamefont {M.}~\bibnamefont {Wiesemann}}, \ and\ \bibinfo {author}
  {\bibfnamefont {G.}~\bibnamefont {Zanderighi}},\ }\href@noop {} {\  (\bibinfo
  {year} {2019})},\ \Eprint {http://arxiv.org/abs/1908.06987} {arXiv:1908.06987
  [hep-ph]} \BibitemShut {NoStop}%
\bibitem [{\citenamefont {Nason}(2004)}]{Nason:2004rx}%
  \BibitemOpen
  \bibfield  {author} {\bibinfo {author} {\bibfnamefont {P.}~\bibnamefont
  {Nason}},\ }\href {\doibase 10.1088/1126-6708/2004/11/040} {\bibfield
  {journal} {\bibinfo  {journal} {JHEP}\ }\textbf {\bibinfo {volume} {11}},\
  \bibinfo {pages} {040} (\bibinfo {year} {2004})},\ \Eprint
  {http://arxiv.org/abs/hep-ph/0409146} {arXiv:hep-ph/0409146 [hep-ph]}
  \BibitemShut {NoStop}%
%%CITATION = HEP-PH/0409146;%%
\bibitem [{\citenamefont {Bahr}\ \emph {et~al.}(2008)\citenamefont {Bahr} \emph
  {et~al.}}]{Bahr:2008pv}%
  \BibitemOpen
  \bibfield  {author} {\bibinfo {author} {\bibfnamefont {M.}~\bibnamefont
  {Bahr}} \emph {et~al.},\ }\href {\doibase 10.1140/epjc/s10052-008-0798-9}
  {\bibfield  {journal} {\bibinfo  {journal} {Eur. Phys. J.}\ }\textbf
  {\bibinfo {volume} {C58}},\ \bibinfo {pages} {639} (\bibinfo {year}
  {2008})},\ \Eprint {http://arxiv.org/abs/0803.0883} {arXiv:0803.0883
  [hep-ph]} \BibitemShut {NoStop}%
%%CITATION = ARXIV:0803.0883;%%
\bibitem [{\citenamefont {Frixione}\ \emph {et~al.}(2007)\citenamefont
  {Frixione}, \citenamefont {Nason},\ and\ \citenamefont
  {Oleari}}]{Frixione:2007vw}%
  \BibitemOpen
  \bibfield  {author} {\bibinfo {author} {\bibfnamefont {S.}~\bibnamefont
  {Frixione}}, \bibinfo {author} {\bibfnamefont {P.}~\bibnamefont {Nason}}, \
  and\ \bibinfo {author} {\bibfnamefont {C.}~\bibnamefont {Oleari}},\ }\href
  {\doibase 10.1088/1126-6708/2007/11/070} {\bibfield  {journal} {\bibinfo
  {journal} {JHEP}\ }\textbf {\bibinfo {volume} {11}},\ \bibinfo {pages} {070}
  (\bibinfo {year} {2007})},\ \Eprint {http://arxiv.org/abs/0709.2092}
  {arXiv:0709.2092 [hep-ph]} \BibitemShut {NoStop}%
%%CITATION = ARXIV:0709.2092;%%
\bibitem [{\citenamefont {Alioli}\ \emph {et~al.}(2010)\citenamefont {Alioli},
  \citenamefont {Nason}, \citenamefont {Oleari},\ and\ \citenamefont
  {Re}}]{Alioli:2010xd}%
  \BibitemOpen
  \bibfield  {author} {\bibinfo {author} {\bibfnamefont {S.}~\bibnamefont
  {Alioli}}, \bibinfo {author} {\bibfnamefont {P.}~\bibnamefont {Nason}},
  \bibinfo {author} {\bibfnamefont {C.}~\bibnamefont {Oleari}}, \ and\ \bibinfo
  {author} {\bibfnamefont {E.}~\bibnamefont {Re}},\ }\href {\doibase
  10.1007/JHEP06(2010)043} {\bibfield  {journal} {\bibinfo  {journal} {JHEP}\
  }\textbf {\bibinfo {volume} {06}},\ \bibinfo {pages} {043} (\bibinfo {year}
  {2010})},\ \Eprint {http://arxiv.org/abs/1002.2581} {arXiv:1002.2581
  [hep-ph]} \BibitemShut {NoStop}%
%%CITATION = ARXIV:1002.2581;%%
\bibitem [{\citenamefont {Grazzini}\ \emph {et~al.}(2018)\citenamefont
  {Grazzini}, \citenamefont {Kallweit},\ and\ \citenamefont
  {Wiesemann}}]{Grazzini:2017mhc}%
  \BibitemOpen
  \bibfield  {author} {\bibinfo {author} {\bibfnamefont {M.}~\bibnamefont
  {Grazzini}}, \bibinfo {author} {\bibfnamefont {S.}~\bibnamefont {Kallweit}},
  \ and\ \bibinfo {author} {\bibfnamefont {M.}~\bibnamefont {Wiesemann}},\
  }\href {\doibase 10.1140/epjc/s10052-018-5771-7} {\bibfield  {journal}
  {\bibinfo  {journal} {Eur. Phys. J.}\ }\textbf {\bibinfo {volume} {C78}},\
  \bibinfo {pages} {537} (\bibinfo {year} {2018})},\ \Eprint
  {http://arxiv.org/abs/1711.06631} {arXiv:1711.06631 [hep-ph]} \BibitemShut
  {NoStop}%
%%CITATION = ARXIV:1711.06631;%%
\bibitem [{\citenamefont {Ball}\ \emph {et~al.}(2017)\citenamefont {Ball} \emph
  {et~al.}}]{Ball:2017nwa}%
  \BibitemOpen
  \bibfield  {author} {\bibinfo {author} {\bibfnamefont {R.~D.}\ \bibnamefont
  {Ball}} \emph {et~al.} (\bibinfo {collaboration} {NNPDF}),\ }\href {\doibase
  10.1140/epjc/s10052-017-5199-5} {\bibfield  {journal} {\bibinfo  {journal}
  {Eur. Phys. J. C}\ }\textbf {\bibinfo {volume} {77}},\ \bibinfo {pages} {663}
  (\bibinfo {year} {2017})},\ \Eprint {http://arxiv.org/abs/1706.00428}
  {arXiv:1706.00428 [hep-ph]} \BibitemShut {NoStop}%
\bibitem [{\citenamefont {Buckley}\ \emph {et~al.}(2015)\citenamefont
  {Buckley}, \citenamefont {Ferrando}, \citenamefont {Lloyd}, \citenamefont
  {Nordström}, \citenamefont {Page}, \citenamefont {Rüfenacht}, \citenamefont
  {Schönherr},\ and\ \citenamefont {Watt}}]{Buckley:2014ana}%
  \BibitemOpen
  \bibfield  {author} {\bibinfo {author} {\bibfnamefont {A.}~\bibnamefont
  {Buckley}}, \bibinfo {author} {\bibfnamefont {J.}~\bibnamefont {Ferrando}},
  \bibinfo {author} {\bibfnamefont {S.}~\bibnamefont {Lloyd}}, \bibinfo
  {author} {\bibfnamefont {K.}~\bibnamefont {Nordström}}, \bibinfo {author}
  {\bibfnamefont {B.}~\bibnamefont {Page}}, \bibinfo {author} {\bibfnamefont
  {M.}~\bibnamefont {Rüfenacht}}, \bibinfo {author} {\bibfnamefont
  {M.}~\bibnamefont {Schönherr}}, \ and\ \bibinfo {author} {\bibfnamefont
  {G.}~\bibnamefont {Watt}},\ }\href {\doibase 10.1140/epjc/s10052-015-3318-8}
  {\bibfield  {journal} {\bibinfo  {journal} {Eur. Phys. J.}\ }\textbf
  {\bibinfo {volume} {C75}},\ \bibinfo {pages} {132} (\bibinfo {year}
  {2015})},\ \Eprint {http://arxiv.org/abs/1412.7420} {arXiv:1412.7420
  [hep-ph]} \BibitemShut {NoStop}%
%%CITATION = ARXIV:1412.7420;%%
\bibitem [{\citenamefont {Salam}\ and\ \citenamefont
  {Rojo}(2009)}]{Salam:2008qg}%
  \BibitemOpen
  \bibfield  {author} {\bibinfo {author} {\bibfnamefont {G.~P.}\ \bibnamefont
  {Salam}}\ and\ \bibinfo {author} {\bibfnamefont {J.}~\bibnamefont {Rojo}},\
  }\href {\doibase 10.1016/j.cpc.2008.08.010} {\bibfield  {journal} {\bibinfo
  {journal} {Comput. Phys. Commun.}\ }\textbf {\bibinfo {volume} {180}},\
  \bibinfo {pages} {120} (\bibinfo {year} {2009})},\ \Eprint
  {http://arxiv.org/abs/0804.3755} {arXiv:0804.3755 [hep-ph]} \BibitemShut
  {NoStop}%
%%CITATION = ARXIV:0804.3755;%%
\bibitem [{\citenamefont {Sjöstrand}\ \emph {et~al.}(2015)\citenamefont
  {Sjöstrand}, \citenamefont {Ask}, \citenamefont {Christiansen},
  \citenamefont {Corke}, \citenamefont {Desai}, \citenamefont {Ilten},
  \citenamefont {Mrenna}, \citenamefont {Prestel}, \citenamefont {Rasmussen},\
  and\ \citenamefont {Skands}}]{Sjostrand:2014zea}%
  \BibitemOpen
  \bibfield  {author} {\bibinfo {author} {\bibfnamefont {T.}~\bibnamefont
  {Sjöstrand}}, \bibinfo {author} {\bibfnamefont {S.}~\bibnamefont {Ask}},
  \bibinfo {author} {\bibfnamefont {J.~R.}\ \bibnamefont {Christiansen}},
  \bibinfo {author} {\bibfnamefont {R.}~\bibnamefont {Corke}}, \bibinfo
  {author} {\bibfnamefont {N.}~\bibnamefont {Desai}}, \bibinfo {author}
  {\bibfnamefont {P.}~\bibnamefont {Ilten}}, \bibinfo {author} {\bibfnamefont
  {S.}~\bibnamefont {Mrenna}}, \bibinfo {author} {\bibfnamefont
  {S.}~\bibnamefont {Prestel}}, \bibinfo {author} {\bibfnamefont {C.~O.}\
  \bibnamefont {Rasmussen}}, \ and\ \bibinfo {author} {\bibfnamefont {P.~Z.}\
  \bibnamefont {Skands}},\ }\href {\doibase 10.1016/j.cpc.2015.01.024}
  {\bibfield  {journal} {\bibinfo  {journal} {Comput. Phys. Commun.}\ }\textbf
  {\bibinfo {volume} {191}},\ \bibinfo {pages} {159} (\bibinfo {year}
  {2015})},\ \Eprint {http://arxiv.org/abs/1410.3012} {arXiv:1410.3012
  [hep-ph]} \BibitemShut {NoStop}%
%%CITATION = ARXIV:1410.3012;%%
\bibitem [{\citenamefont {Cabouat}\ and\ \citenamefont
  {Sjöstrand}(2018)}]{Cabouat:2017rzi}%
  \BibitemOpen
  \bibfield  {author} {\bibinfo {author} {\bibfnamefont {B.}~\bibnamefont
  {Cabouat}}\ and\ \bibinfo {author} {\bibfnamefont {T.}~\bibnamefont
  {Sjöstrand}},\ }\href {\doibase 10.1140/epjc/s10052-018-5645-z} {\bibfield
  {journal} {\bibinfo  {journal} {Eur. Phys. J. C}\ }\textbf {\bibinfo {volume}
  {78}},\ \bibinfo {pages} {226} (\bibinfo {year} {2018})},\ \Eprint
  {http://arxiv.org/abs/1710.00391} {arXiv:1710.00391 [hep-ph]} \BibitemShut
  {NoStop}%
\bibitem [{\citenamefont {Nagy}\ and\ \citenamefont
  {Soper}(2010)}]{Nagy:2009vg}%
  \BibitemOpen
  \bibfield  {author} {\bibinfo {author} {\bibfnamefont {Z.}~\bibnamefont
  {Nagy}}\ and\ \bibinfo {author} {\bibfnamefont {D.~E.}\ \bibnamefont
  {Soper}},\ }\href {\doibase 10.1007/JHEP03(2010)097} {\bibfield  {journal}
  {\bibinfo  {journal} {JHEP}\ }\textbf {\bibinfo {volume} {03}},\ \bibinfo
  {pages} {097} (\bibinfo {year} {2010})},\ \Eprint
  {http://arxiv.org/abs/0912.4534} {arXiv:0912.4534 [hep-ph]} \BibitemShut
  {NoStop}%
\bibitem [{\citenamefont {Dasgupta}\ \emph {et~al.}(2018)\citenamefont
  {Dasgupta}, \citenamefont {Dreyer}, \citenamefont {Hamilton}, \citenamefont
  {Monni},\ and\ \citenamefont {Salam}}]{Dasgupta:2018nvj}%
  \BibitemOpen
  \bibfield  {author} {\bibinfo {author} {\bibfnamefont {M.}~\bibnamefont
  {Dasgupta}}, \bibinfo {author} {\bibfnamefont {F.~A.}\ \bibnamefont
  {Dreyer}}, \bibinfo {author} {\bibfnamefont {K.}~\bibnamefont {Hamilton}},
  \bibinfo {author} {\bibfnamefont {P.~F.}\ \bibnamefont {Monni}}, \ and\
  \bibinfo {author} {\bibfnamefont {G.~P.}\ \bibnamefont {Salam}},\ }\href
  {\doibase 10.1007/JHEP09(2018)033} {\bibfield  {journal} {\bibinfo  {journal}
  {JHEP}\ }\textbf {\bibinfo {volume} {09}},\ \bibinfo {pages} {033} (\bibinfo
  {year} {2018})},\ \Eprint {http://arxiv.org/abs/1805.09327} {arXiv:1805.09327
  [hep-ph]} \BibitemShut {NoStop}%
%%CITATION = ARXIV:1805.09327;%%
\bibitem [{\citenamefont {Bewick}\ \emph {et~al.}(2020)\citenamefont {Bewick},
  \citenamefont {Ferrario~Ravasio}, \citenamefont {Richardson},\ and\
  \citenamefont {Seymour}}]{Bewick:2019rbu}%
  \BibitemOpen
  \bibfield  {author} {\bibinfo {author} {\bibfnamefont {G.}~\bibnamefont
  {Bewick}}, \bibinfo {author} {\bibfnamefont {S.}~\bibnamefont
  {Ferrario~Ravasio}}, \bibinfo {author} {\bibfnamefont {P.}~\bibnamefont
  {Richardson}}, \ and\ \bibinfo {author} {\bibfnamefont {M.~H.}\ \bibnamefont
  {Seymour}},\ }\href {\doibase 10.1007/JHEP04(2020)019} {\bibfield  {journal}
  {\bibinfo  {journal} {JHEP}\ }\textbf {\bibinfo {volume} {04}},\ \bibinfo
  {pages} {019} (\bibinfo {year} {2020})},\ \Eprint
  {http://arxiv.org/abs/1904.11866} {arXiv:1904.11866 [hep-ph]} \BibitemShut
  {NoStop}%
\bibitem [{\citenamefont {Dasgupta}\ \emph {et~al.}(2020)\citenamefont
  {Dasgupta}, \citenamefont {Dreyer}, \citenamefont {Hamilton}, \citenamefont
  {Monni}, \citenamefont {Salam},\ and\ \citenamefont
  {Soyez}}]{Dasgupta:2020fwr}%
  \BibitemOpen
  \bibfield  {author} {\bibinfo {author} {\bibfnamefont {M.}~\bibnamefont
  {Dasgupta}}, \bibinfo {author} {\bibfnamefont {F.~A.}\ \bibnamefont
  {Dreyer}}, \bibinfo {author} {\bibfnamefont {K.}~\bibnamefont {Hamilton}},
  \bibinfo {author} {\bibfnamefont {P.~F.}\ \bibnamefont {Monni}}, \bibinfo
  {author} {\bibfnamefont {G.~P.}\ \bibnamefont {Salam}}, \ and\ \bibinfo
  {author} {\bibfnamefont {G.}~\bibnamefont {Soyez}},\ }\href@noop {} {\
  (\bibinfo {year} {2020})},\ \Eprint {http://arxiv.org/abs/2002.11114}
  {arXiv:2002.11114 [hep-ph]} \BibitemShut {NoStop}%
\bibitem [{\citenamefont {Forshaw}\ \emph {et~al.}(2020)\citenamefont
  {Forshaw}, \citenamefont {Holguin},\ and\ \citenamefont
  {Plätzer}}]{Forshaw:2020wrq}%
  \BibitemOpen
  \bibfield  {author} {\bibinfo {author} {\bibfnamefont {J.~R.}\ \bibnamefont
  {Forshaw}}, \bibinfo {author} {\bibfnamefont {J.}~\bibnamefont {Holguin}}, \
  and\ \bibinfo {author} {\bibfnamefont {S.}~\bibnamefont {Plätzer}},\
  }\href@noop {} {\  (\bibinfo {year} {2020})},\ \Eprint
  {http://arxiv.org/abs/2003.06400} {arXiv:2003.06400 [hep-ph]} \BibitemShut
  {NoStop}%
\bibitem [{\citenamefont {Butterworth}\ \emph {et~al.}(2016)\citenamefont
  {Butterworth} \emph {et~al.}}]{Butterworth:2015oua}%
  \BibitemOpen
  \bibfield  {author} {\bibinfo {author} {\bibfnamefont {J.}~\bibnamefont
  {Butterworth}} \emph {et~al.},\ }\href {\doibase
  10.1088/0954-3899/43/2/023001} {\bibfield  {journal} {\bibinfo  {journal} {J.
  Phys.}\ }\textbf {\bibinfo {volume} {G43}},\ \bibinfo {pages} {023001}
  (\bibinfo {year} {2016})},\ \Eprint {http://arxiv.org/abs/1510.03865}
  {arXiv:1510.03865 [hep-ph]} \BibitemShut {NoStop}%
%%CITATION = ARXIV:1510.03865;%%
\bibitem [{\citenamefont {Ball}\ \emph {et~al.}(2015)\citenamefont {Ball} \emph
  {et~al.}}]{Ball:2014uwa}%
  \BibitemOpen
  \bibfield  {author} {\bibinfo {author} {\bibfnamefont {R.~D.}\ \bibnamefont
  {Ball}} \emph {et~al.} (\bibinfo {collaboration} {NNPDF}),\ }\href {\doibase
  10.1007/JHEP04(2015)040} {\bibfield  {journal} {\bibinfo  {journal} {JHEP}\
  }\textbf {\bibinfo {volume} {04}},\ \bibinfo {pages} {040} (\bibinfo {year}
  {2015})},\ \Eprint {http://arxiv.org/abs/1410.8849} {arXiv:1410.8849
  [hep-ph]} \BibitemShut {NoStop}%
%%CITATION = ARXIV:1410.8849;%%
\bibitem [{\citenamefont {Harland-Lang}\ \emph {et~al.}(2015)\citenamefont
  {Harland-Lang}, \citenamefont {Martin}, \citenamefont {Motylinski},\ and\
  \citenamefont {Thorne}}]{Harland-Lang:2014zoa}%
  \BibitemOpen
  \bibfield  {author} {\bibinfo {author} {\bibfnamefont {L.~A.}\ \bibnamefont
  {Harland-Lang}}, \bibinfo {author} {\bibfnamefont {A.~D.}\ \bibnamefont
  {Martin}}, \bibinfo {author} {\bibfnamefont {P.}~\bibnamefont {Motylinski}},
  \ and\ \bibinfo {author} {\bibfnamefont {R.~S.}\ \bibnamefont {Thorne}},\
  }\href {\doibase 10.1140/epjc/s10052-015-3397-6} {\bibfield  {journal}
  {\bibinfo  {journal} {Eur. Phys. J.}\ }\textbf {\bibinfo {volume} {C75}},\
  \bibinfo {pages} {204} (\bibinfo {year} {2015})},\ \Eprint
  {http://arxiv.org/abs/1412.3989} {arXiv:1412.3989 [hep-ph]} \BibitemShut
  {NoStop}%
%%CITATION = ARXIV:1412.3989;%%
\bibitem [{\citenamefont {Dulat}\ \emph {et~al.}(2016)\citenamefont {Dulat},
  \citenamefont {Hou}, \citenamefont {Gao}, \citenamefont {Guzzi},
  \citenamefont {Huston}, \citenamefont {Nadolsky}, \citenamefont {Pumplin},
  \citenamefont {Schmidt}, \citenamefont {Stump},\ and\ \citenamefont
  {Yuan}}]{Dulat:2015mca}%
  \BibitemOpen
  \bibfield  {author} {\bibinfo {author} {\bibfnamefont {S.}~\bibnamefont
  {Dulat}}, \bibinfo {author} {\bibfnamefont {T.-J.}\ \bibnamefont {Hou}},
  \bibinfo {author} {\bibfnamefont {J.}~\bibnamefont {Gao}}, \bibinfo {author}
  {\bibfnamefont {M.}~\bibnamefont {Guzzi}}, \bibinfo {author} {\bibfnamefont
  {J.}~\bibnamefont {Huston}}, \bibinfo {author} {\bibfnamefont
  {P.}~\bibnamefont {Nadolsky}}, \bibinfo {author} {\bibfnamefont
  {J.}~\bibnamefont {Pumplin}}, \bibinfo {author} {\bibfnamefont
  {C.}~\bibnamefont {Schmidt}}, \bibinfo {author} {\bibfnamefont
  {D.}~\bibnamefont {Stump}}, \ and\ \bibinfo {author} {\bibfnamefont {C.~P.}\
  \bibnamefont {Yuan}},\ }\href {\doibase 10.1103/PhysRevD.93.033006}
  {\bibfield  {journal} {\bibinfo  {journal} {Phys. Rev.}\ }\textbf {\bibinfo
  {volume} {D93}},\ \bibinfo {pages} {033006} (\bibinfo {year} {2016})},\
  \Eprint {http://arxiv.org/abs/1506.07443} {arXiv:1506.07443 [hep-ph]}
  \BibitemShut {NoStop}%
%%CITATION = ARXIV:1506.07443;%%
\bibitem [{\citenamefont {Nason}\ and\ \citenamefont
  {Oleari}(2013)}]{Nason:2013uba}%
  \BibitemOpen
  \bibfield  {author} {\bibinfo {author} {\bibfnamefont {P.}~\bibnamefont
  {Nason}}\ and\ \bibinfo {author} {\bibfnamefont {C.}~\bibnamefont {Oleari}},\
  }\href@noop {} {\  (\bibinfo {year} {2013})},\ \Eprint
  {http://arxiv.org/abs/1303.3922} {arXiv:1303.3922 [hep-ph]} \BibitemShut
  {NoStop}%
\bibitem [{\citenamefont {Catani}\ and\ \citenamefont
  {Webber}(1997)}]{Catani:1997xc}%
  \BibitemOpen
  \bibfield  {author} {\bibinfo {author} {\bibfnamefont {S.}~\bibnamefont
  {Catani}}\ and\ \bibinfo {author} {\bibfnamefont {B.}~\bibnamefont
  {Webber}},\ }\href {\doibase 10.1088/1126-6708/1997/10/005} {\bibfield
  {journal} {\bibinfo  {journal} {JHEP}\ }\textbf {\bibinfo {volume} {10}},\
  \bibinfo {pages} {005} (\bibinfo {year} {1997})},\ \Eprint
  {http://arxiv.org/abs/hep-ph/9710333} {arXiv:hep-ph/9710333} \BibitemShut
  {NoStop}%
\end{thebibliography}%
\end{document}